\documentclass[12pt]{article}
\usepackage{amsmath}
\usepackage{amssymb}
\usepackage{amsfonts}
\usepackage{epsfig}
\usepackage{mathrsfs}
\usepackage{amsmath,amsthm,amssymb,amscd}
\usepackage{MnSymbol}
\usepackage{rotating}
\usepackage{multirow}
\usepackage{float}
\usepackage{hyperref}
\usepackage[english]{babel}
\usepackage{floatpag}
\usepackage{enumerate}
\usepackage{threeparttable}
\usepackage{algorithm}
\usepackage{algorithmic}
\usepackage{xcolor} % for colored text

\newcommand\appendix@section[1]{\refstepcounter{section}\orig@section*{#1}\addcontentsline{toc}{section}{#1}}
\let\orig@section\section\g@addto@macro\appendix{\let\section\appendix@section}
\makeatother \makeatletter
\renewcommand\footnoterule{\kern-3\p@ \hrule width 1\columnwidth \kern 2.6\p@}%Footnote Line
\newskip\@footindent
\renewcommand\@footindent{0pt}%Footnote indent
\@addtoreset{footnote}{page}
\long\def\@makefntext#1{\@setpar{\@@par\@tempdima \hsize
\advance\@tempdima-\@footindent \parshape \@ne \@footindent
\@tempdima}\par \noindent \hbox to
\z@{\hss\@thefnmark£º\hspace{0.2em}}#1}

\def\@makefnmark{\hbox{\textsuperscript{\@thefnmark}}}
\makeatother

\newtheorem{pro}{Proposition}
\newtheorem{thm}{Theorem}
\newtheorem{cor}{Corollary}
\newtheorem{lem}{Lemma}[section]

\title{Surprise sampling: improving and extending the local case-control sampling}
\author{ \vspace{2mm} Xinwei Shen \ \ \ Kani Chen\\
 Department of Mathematics\\
 Hong Kong University of Science and Technology\\
 \vspace{10mm}
 Clear Water Bay, Kowloon, Hong Kong\\
 \and
 \vspace{2mm} Wen Yu$^\ast$\\
 Department of Statistics \\
 School of Management\\
 Fudan University\\
  \vspace{10mm}
 Shanghai 200433, P.R.China\\
{\small $^\ast$Corresponding author: \ wenyu@fudan.edu.cn}
 }
\date{ }
\begin{document}
\maketitle
\newpage
\begin{abstract}
Fithian and Hastie (2014) proposed a new sampling scheme called local case-control (LCC) sampling that achieves stability and efficiency by utilizing a clever adjustment pertained to the logistic model. It is particularly useful for classification with large and imbalanced data. This paper proposes a more general sampling scheme based on a working principle that data points deserve higher sampling probability if they contain more information or appear ``surprising" in the sense of, for example, a large error of pilot prediction or a large absolute score. Compared with the relevant existing sampling schemes, as reported in Fithian and Hastie  (2014) and Ai, et al. (2018), the proposed one has several advantages. It adaptively gives out the optimal forms to a variety of objectives, including the LCC and Ai et al. (2018)'s sampling as special cases. Under same model specifications, the proposed estimator also performs no worse than those in the literature. The estimation procedure is valid even if the model is misspecified and/or the pilot estimator is inconsistent or dependent on full data. We present theoretical justifications of the claimed advantages and optimality of the estimation and the sampling design. Different from Ai, et al. (2018), our large sample theory are population-wise rather than data-wise. Moreover, the proposed approach can be applied to unsupervised learning studies, since it essentially only requires a specific loss function and no response-covariate structure of data is needed. Numerical studies are carried out and the evidence in support of the theory is shown.
\end{abstract}

\vfill \hrule \vskip 6pt \noindent {\em MSC:}  \ 62D05, 62J12\\
\noindent {\em Some key words}: Generalized linear models, Horvitz-Thompson estimator, Local case-control sampling, Model mis-specification, Subsampling.

\newpage

\section{Introduction}

Nowadays, with the rapid development of data capturing and storage techniques, people meet huge amounts of data in various fields. With the growth of the data size, computational capacity becomes more and more crucial to implement efficient data analysis. Despite significant progress on computer hardware, computational ability may still become a major constraint for data analysis when the data size is sufficiently large. For instance, it might be very time and resource consuming if we want to try a variety of competing models instead of only fitting one or two predetermined models, or if we need to refit the model from time to time with new observations arriving continuously, or if we need to apply the data partition and reusing techniques such as cross-validation, bootstrapping, bagging, and so on. All these computationally intensive procedures require tremendous computational costs, especially for those large data sets.

A simple approach to reduce the computational cost is to draw a subsample from the full data set and then analyze the subsample. The easiest subsampling design is to draw the data uniformly. However, uniform subsampling might be very inefficient for some data structures. For instance, in classification problems with two classes (one class for positive examples called ``cases" and the other for negative examples called ``controls"), when the classes are imbalanced, that is, one of the classes (usually the class of cases) is rare and the other is dominant, uniform subsampling is unfavorable because it ignores the unequal importance of the data points. For imbalanced data, case-control sampling is a well-known subsampling design (Mantel and Haenszel, 1959; Miettinen, 1976; Breslow, Day et al., 1980). It draws uniform subsamples from each of the two classes with different sampling percentages. Often comparable number of cases and controls are sampled, yielding a subsample with no obvious imbalance, to increase the efficiency of estimation. Anderson (1972) and Prentice and Pyke (1979) showed that by fitting a logistic model on the case-control subsample and then making a simple adjustment, one can still get a consistent estimator for the regression parameters if the logistic model is correctly specified. Thus, case-control design can help reduce the computational burden and retain satisfactory estimation efficiency under the imbalanced structure. Besides classification, imbalanced data structure also appears for other data types. A survey on predictive modeling under various imbalanced distributions is provided by Branco et al. (2015).

Subsampling designs have also received attention in epidemiological cohort studies. The cohorts to follow up usually involve a great number of subjects and the occurrence of a certain disease is interested. In most cases, the occurrence rate of the disease is low during the entire follow-up time, so the cohort is essentially imbalanced. Meanwhile, the collection of covariate information on all involved subjects can be very expensive and time consuming because of the large cohort size. Hence, subsampling designs were developed to save sampling cost and time. The widely studied ones include case-cohort design (Prentice, 1986) and nested case-control design (Thomas, 1977). Case-control design can also be used in cohort studies. More generalized case-cohort designs aiming to improve the estimation efficiency were developed later by Chen and Lo (1999), Chen (2001), and more recently, Yao et al. (2017). The sampling probability of all these subsampling designs depends on the observed follow-up times and the censoring indicators, but not the covariates. These designs are usually used for covariates ascertainment, that is, the covariates are observed only when the subject is selected into the subsample.

Statistical models are usually imposed for data analysis, but in real applications the models are easily misspecified. For some machine learning approaches, models are just used to derive meaningful loss functions, without caring about if they are correctly specified or not. Under these circumstances, the target parameter becomes a certain population risk minimizer corresponding to the loss function used (Huber, 2011). When there exists possibly model mis-specification, robust analysis procedures are often preferred. However, as Fithian and Hastie (2014) pointed out, in classification problems, when the logistic model is misspecified, the standard case-control sampling can not provide consistent estimator of the target parameter. To overcome this drawback, Fithian and Hastie (2014) proposed a local case-control (LCC) sampling design, which depends not only on the class label, but also on the predictors and a pilot estimate (a pilot estimate is a ``good" guess for the target parameter). Their design provides a clever way to remedy imbalance locally throughout the feature space and possesses potential robustness against model mis-specification. Moreover, an elegant LCC estimate mimicking the full sample maximum likelihood estimate (MLE) is proposed. They showed that when the logistic model is correctly specified and the pilot is consistent and independent of the full data, the asymptotic variance of their proposed estimate is twice the variance of the full data MLE; when the logistic model is misspecified, the proposed estimate is still consistent and asymptotically normal as long as the pilot is consistent and independent of the data.

The LCC design aims to simultaneously speed up computation and provide a simple procedure to obtain a good estimate even under model mis-specification, but Fithian and Hastie (2014) did not discuss the optimality of the design. More recently, Wang et al. (2018) modified the LCC design smartly to get an optimal subsampling method that minimizes the conditional asymptotic mean squared error (MSE) of the subsample estimator given the full data. Later on, Ai et al. (2018) extended the method to generalized linear models with canonical links. However, they only considered the optimization criterion regarding the conditional MSE. Some other non-uniform subsampling designs for the linear model include Ma et al. (2015) and Wang et al. (2018). We propose an improved subsampling design which accommodates various types of statistical learning objectives and includes the LCC sampling and Ai et al. (2018)'s sampling as special cases. For estimating the target parameter based on the subsample, we apply the Horvitz-Thompson (HT) type estimation (Horvitz and Thompson, 1952). The new sampling design is derived by optimizing certain well-defined criteria, such as prediction accuracy, estimation accuracy, or MSE. For different criteria, the proposed design has its corresponding form adaptively. Basically, it draws a data point with a large error of pilot prediction or a large score into the subsample with higher probability. Provided a pilot estimate, such a data point is in certain sense unusual, or ``surprising", for that given pilot, so we call the proposed design surprise sampling design.

The advantages of the proposed surprise sampling are summarized as follows.
\begin{itemize}
\item [] (i) The surprise sampling design is derived based on well-defined objectives. For a specific objective, the corresponding design is optimal. Meanwhile, the proposed design flexibly adapts varying objectives. In contrast, the objective of the LCC design is not clearly defined and the optimality is not discussed in Fithian and Hastie (2014). In Wang et al. (2018) and Ai et al. (2018), only the objective of minimizing the conditional MSE is considered.
\item [] (ii) The proposed estimators are always consistent and asymptotically normal, regardless of the correctness of the model specification and the consistency of the pilot estimate. The consistency of Fithian and Hastie (2014)'s estimator does not require the logistic model specification to be correct, but needs the consistency of the pilot. The pilots used by Wang et al. (2018) and Ai et al. (2018) are also consistent.
\item [] (iii) If the pilot estimate is consistent and the logistic model is correctly specified, the proposed estimator is no worse than Fithian and Hastie (2014)'s estimator in the sense that they have the same asymptotic efficiency under the LCC sampling.
\item [] (iv) The validity of the proposed estimation procedure does not require the pilot estimate to be independent of the full data, while Fithian and Hastie (2014)'s estimator requires the independence. This relaxation is useful in applications especially when there is no other data source and one needs to get the pilot from the full data.
\item [] (v) The large sample properties derived for the proposed estimators are population-wise (unconditional), while the parallel large sample properties in Wang et al. (2018) and Ai et al. (2018) are developed data-wise (conditional on the full data).
\item [] (vi) The proposed approach can be generally applied to not only supervised learning such as classification and regression but also unsupervised learning tasks, because it essentially only requires a well-defined loss function with a finite-dimensional parameter. Hence the application of the approach is more than the scope of Logistic model or the other generalized linear models.
\end{itemize}

We proceed as follows. Section 2 introduces the notation and describes the problem setting. In Section 3, we present the main idea of the proposed surprise sampling and the specific forms of the sampling design are respectively derived to reach various objectives, such as best prediction accuracy and estimation accuracy. Section 4 gives out the large sample properties of the HT type estimator under the surprise sampling design. In Section 5, extensive simulation studies are carried out to show the effectiveness of the proposed approach. The results of real data analysis are provided in Section 6. Section 7 concludes. All the technique details are summarized in the Appendix.

\section{Notation and problem setting}

Suppose the full data consists $n$ subjects and let $d_i$ be the observed data point for the $i$-th subject. We assume that $d_i$, $i=1,\ldots,n$, are independent and identically distributed (i.i.d.) copies of a random element $D$ whose distribution stands for the population. For supervised learning problems, $D$ can be decomposed into a response, denoted by $Y$, and a $q$-dimensional predictor or covariate vector, denoted by $X$. Correspondingly, the observed data points $d_i=(y_i,x_i)$, $i=1,\ldots,n$, are i.i.d. copies of $D=(Y,X)$. We aim to predict $Y$ through $X$, or learn the regression function $f(x)=\mathsf{E}(Y|X=x)$, based on the observed data. Usually, a model is imposed on $f(x)$ characterized by a $p$-dimensional parameter, denoted by $\theta$ throughout the paper, and a corresponding loss function is minimized to obtain an estimate of $\theta$. Write the loss function as $l(d;\theta)$, where $d=(y,x)$. The loss function can take the form of squared loss, negative log-likelihood, hinge loss, Huber loss, etc. Some classical choices include Logistic regression with $l(d;\theta)=-y(\alpha+\beta^\top x)+\log(1+\exp(\alpha+\beta^\top x))$ for a binary response, where $\theta=(\alpha,\beta^\top)$, Poisson log-linear model with $l(d;\theta)=-y(\alpha+\beta^\top x)+\exp(\alpha+\beta^\top x)$ for a counting response, linear model with squared loss $l(d;\theta)=(y-\alpha-\beta^\top x)^2/\sigma^2$ for a continuous response, where $\theta=(\alpha,\beta^\top,\sigma)$, and also nonlinear models such as neural networks with squared loss or cross-entropy loss. For unsupervised learning tasks, there is no response-covariate structure of data. Typically, we derive a loss function $l(d;\theta)$ based on a certain purpose and aim at minimizing the loss with respect to $\theta$. For instance, in the geometric view of principal component analysis (PCA), the goal is to find a $k$-dimensional affine space in $\mathbb{R}^q$ that best approximate the $n$ examples in terms of Euclidean distance. Parameterizing the affine space by $\alpha+U\beta$ where $U$ consists of $k$-columns of an orthogonal basis of the space, we end up with the following loss function $l(d;\theta)=\|d-(\alpha+U\beta)\|^2$, where $\theta=\{\alpha,U,\beta\}$ and $\|\cdot\|$ stands for the Euclidean norm.

The target parameter that we aim to estimate is the so-called population risk minimizer which minimizes the population risk $R(\theta)=\mathsf{E}[l(D;\theta)]$, that is,
\begin{eqnarray*}
\theta^\ast=\mbox{arg min}_\theta R(\theta).
\end{eqnarray*}
In supervised learning, when the regression model is correctly specified, that is, there exists some $\theta_0$ satisfying $f(x)=f_{\theta_0}(x)$, then it is easy to see that $\theta_0=\theta^\ast$. However, as we mention in Section 1, in many real applications, $f(x)$ does not satisfy the specified model and this is the so-called model mis-specification. Meanwhile, in some unsupervised learning tasks such as PCA, there is no imposed model. It is well-known that no matter the model specification holds or not, ${\theta}^\ast$ can be well defined under general conditions. The full sample version of the risk minimizer, denoted by $\hat{\theta}^\ast$,  is given by $\hat{\theta}^\ast=\mbox{arg min}_\theta\sum_{i=1}^n l(d_i;\theta)$. Under suitable regularity conditions, one can show that $\hat{\theta}^\ast$ is consistent for $\theta^\ast$, that is, $\hat{\theta}^\ast$ converges in probability to $\theta^\ast$ (Huber, 2011).

When $n$ is sufficiently large, subsampling designs are preferred to save computational cost. For each $i=1,\ldots,n$, let $\Delta_i$ be a 0-1 valued binary indicator, indicating whether the $i$th data point is sampled into the subsample ($\Delta_i=1$) or not ($\Delta_i=0$). Let $\tilde{\theta}$ be a pilot estimate (i.e., a guess of $\theta^\ast$). Let $\pi_i$ be the conditional sampling probability of the $i$th data point, i.e., the conditional probability of $\Delta_i=1$, given all the observed data and the pilot estimate. Thus, $\pi_i$ may depend on the observed data and the pilot. Also, $\Delta_i$'s are generated independently with each other given the observed data and the pilot.

For binary response, the LCC design proposed by Fithian and Hastie (2014) sets $\pi_i=|y_i-p(\tilde{\alpha}+\tilde{\beta}^\top x_i)|$, where $p(t)=\exp(t)/[1+\exp(t)]$ and $(\tilde{\alpha},\tilde{\beta})$ is a pilot estimate of $(\alpha,\beta)$. After obtaining the subsample, they fit a Logistic regression to the subsample and then do a simple adjustment to get the estimator of the target parameter. The main advantage of their estimation procedure is that it basically maintains the original form of the maximum likelihood estimation with full data. However, as we mention in Section 1, the validity of their approach heavily relies on the consistency of the pilot estimate, yet it has not been extended to other regression models or more general learning tasks. We apply the HT type estimation. Specifically, the HT type estimator is given by
\begin{eqnarray}\label{2.1}
\hat{\theta}=\mbox{arg min}_\theta\sum_{i=1}^n\frac{\Delta_i}{\pi_i}l(d_i;\theta).
\end{eqnarray}
The HT type estimation is general enough for various kinds of loss functions. The computational complexity of (\ref{2.1}) is similar to that of full data. For instance, if the objective function based on the full data is convex, so is the objective function in (\ref{2.1}). Meanwhile, by inverting the sampling probability, it is quite intuitive to expect the consistency of the HT type estimator regardless of the consistency of the pilot estimate. More importantly, based on the HT type estimation, we can derive an optimal form of the subsampling design for a specific objective. The details of the derivation is discussed in the following section.

Before proceeding to the next section, we introduce some more necessary notation. Let $g(d;\theta)=\partial l(d;\theta)/\partial\theta$. When $l$ is the negative log-likelihood function, $g$ becomes the score vector. Let $G(d;\theta)=\partial^2 l(d;\theta)/\partial\theta\partial\theta^\top$ and $A=\mathsf{E}[G(D;\theta^\ast)]$. For any column vector $a$, $a^{\otimes2}$ stands for $aa^\top$. We use $\|\cdot\|$ to stand for the Euclidean norm.

\section{Surprise sampling design}

To give out the specific form of the proposed subsampling design, we first heuristically present the asymptotic properties of the HT type estimator $\hat{\theta}$. By the definition of $\hat{\theta}$, it is easy to see that $\sum_{i=1}^n\Delta_ig(d_i;\hat{\theta})/\pi_i=0$. Under suitable conditions, we can show that $\hat{\theta}$ is consistent for $\theta^\ast$ and that
\begin{eqnarray}\label{3.0}
\sqrt{n}(\hat{\theta}-\theta^\ast)=-A^{-1}\frac{1}{\sqrt{n}}\sum_{i=1}^n\frac{\Delta_i}{\pi_i}g(d_i;\theta^\ast)+o_p(1).
\end{eqnarray}
Furthermore, we can show that $\sqrt{n}(\hat{\theta}-\theta^\ast)$ converges in distribution to a Gaussian vector, denoted by $Z$, with mean zero and variance-covariance matrix $A^{-1}V_{\pi}A^{-1}$, where $V_{\pi}=\mathsf{E}[g(D;\theta^\ast)^{\otimes 2}/\pi]$ and $\pi$ is a probability that may depend on $D$. We give out sufficient conditions that guarantee the above results in Appendix A.2.

\subsection{Overall prediction accuracy}

If the main purpose of the data analysis is prediction, the minimized population prediction error $R(\theta^\ast)=\mathsf{E}[l(D;\theta^\ast)]$ is a natural criterion to measure the prediction accuracy. Based on a subsampling percentage $\pi$ and the corresponding HT type estimator $\hat{\theta}$, the prediction error can be measured by $R(\hat{\theta})$, which is $R(\theta)$ evaluated at $\theta=\hat{\theta}$. By Taylor expansion, we have that
\begin{eqnarray}
nR(\hat{\theta})-nR(\theta^\ast)&=&n\mathsf{E}[g(D;\theta^\ast)](\hat{\theta}-\theta^\ast)\nonumber\\
& &+\frac{n}{2}(\hat{\theta}-\theta^\ast)^\top\mathsf{E}[G(D;\theta^\ast)](\hat{\theta}-\theta^\ast)+o_p(1)\nonumber\\
&=&\frac{n}{2}(\hat{\theta}-\theta^\ast)^\top A(\hat{\theta}-\theta^\ast)+o_p(1).\label{3.1}
\end{eqnarray}
The leading term in (\ref{3.1}) converges in distribution to $Z^\top A Z/2$ according to the asymptotic property of $\hat{\theta}$. By some calculation, we have that
\begin{eqnarray}
\mathsf{E}(Z^\top A Z)&=&\mathsf{tr}[\mathsf{E}(Z^\top A Z)]=\mathsf{tr}[A\mathsf{E}(ZZ^\top)]=\mathsf{tr}(V_\pi A^{-1})\nonumber\\
&=&\mathsf{E}\left[\mathsf{tr}\left(\frac{A^{-1/2}g(D;\theta^\ast)^{\otimes 2}A^{-1/2}}{\pi}\right)\right]=\mathsf{E}\left(\frac{\|A^{-1/2}g(D;\theta^\ast)\|^2}{\pi}\right).\label{3.2}
\end{eqnarray}
Then it is natural to select $\pi$ that minimizes (\ref{3.2}), which can be treated as the average difference between the prediction error based on the subsample and the minimized population prediction error. The following proposition, proved in Appendix A.1, gives out the optimal form of $\pi$.

\begin{pro}\label{pro1}
Let $r\in(0,1)$ be a constant. The optimal $\pi$ to minimize (\ref{3.2}), subject to $\mathsf{E}(\pi)\leqslant r$, is given by
\begin{eqnarray*}
\pi=\left(c\|A^{-1/2}g(D;\theta^\ast)\|\right)\wedge1,
\end{eqnarray*}
where $c$ is the largest constant such that $\mathsf{E}(\pi)\leqslant r$ and for constants $a$ and $b$, $a\wedge b=\min\{a,b\}$.
\end{pro}

The constant $r$ is used to control the subsampling rate. Note that the optimal $\pi$ depends on some unknown quantities, such as $A$ and $\theta^\ast$. Given a pilot estimate $\tilde{\theta}$, we propose the following subsampling design
\begin{eqnarray}\label{3.3}
\pi_i=\left(c\left\|\tilde{A}^{-1/2}g(d_i;\tilde{\theta})\right\|\right)\wedge1,
\end{eqnarray}
$i=1,\ldots,n$, where $\tilde{A}=n^{-1}\sum_{i=1}^nG(d_i;\tilde{\theta})$ and $c$ is a constant selected to reach the predetermined subsampling rate. For the given subsampling rate $r$, we design an algorithm based on the bisection method to get the largest $c$ such that $n^{-1}\sum_{i=1}^n\pi_i\leqslant r$.

The proposed subsampling probability is proportional to the score evaluated at the pilot. When the value is relatively large for a data point, it implies that this data point has a large prediction error for the given pilot estimate, that is, it is somewhat more ``surprising" than the ones with smaller score values. Under such design, the data point with a larger score value is more likely to be drawn into the subsample. Thus, we call the proposed subsampling design ``surprise" sampling, or score sampling.

\subsection{Estimation accuracy}

If the main concern is the estimation accuracy of $v^\top\hat{\theta}$, as an estimator of $v^\top\theta^\ast$, where $v$ is a given $p$-dimensional constant vector, then the objective becomes to give out an efficient subsampling scheme to increase the estimation accuracy. Without loss of generality, assume that $\|v\|=1$. From (\ref{3.0}), we have that $\sqrt{n}v^\top(\hat{\theta}-\theta^\ast)$ converges in distribution to a Gaussian vector with mean zero and variance $v^\top A^{-1}V_\pi A^{-1}v$. Then we select $\pi$ to minimize this asymptotic variance.

\begin{pro}\label{pro2}
Let $r\in(0,1)$ be a constant. The optimal $\pi$ to minimize $v^\top A^{-1}V_\pi A^{-1}v$, subject to $\mathsf{E}(\pi)\leqslant r$, is given by
\begin{eqnarray*}
\pi=\left(c\left|v^\top A^{-1}g(D;\theta^\ast)\right|\right)\wedge1,
\end{eqnarray*}
where $c$ is the largest constant such that $\mathsf{E}(\pi)\leqslant r$.
\end{pro}

\noindent Given a pilot $\tilde{\theta}$, the proposed subsampling design is given by
\begin{eqnarray}\label{3.4}
\pi_i=\left(c\left|v^\top\tilde{A}^{-1}g(d_i;\tilde{\theta})\right|\right)\wedge1,
\end{eqnarray}
$i=1,\ldots,n$, where $c$ is a constant selected to reach the predetermined subsampling rate. Again, the proposed design is proportional to the score evaluated at $\tilde{\theta}$.

To illustrate the relationship between the proposed surprise sampling and the LCC sampling, we discuss the case of generalized linear models with $D=(Y,X)$ and $d_i=(y_i,x_i)$. Here, let $Z=(1,X^\top)^\top$ and $\theta$ be the collection of all the regression parameters (i.e., the coefficients corresponding to $X$ plus the intercept). For a generalized linear model, the conditional probability or density function of $Y$ given $X=x$, denoted by $p(y|x;\theta)$, can be written as the form of $\psi(y,\theta^\top z)$, where $\psi$ is a known function up to a finite-dimensional parameter and $z=(1,x^\top)^\top$. A natural choice of the loss function is the negative log likelihood $l(y,x;\theta)=-\log\psi(y,\theta^\top z)$. Define $S(y,t)=\frac{\partial}{\partial t}\psi(y,t)/\psi(y,t)$. Then it is easy to see that $g(y,x;\theta)=-S(y,\theta^\top z)z$. When the model is correctly specified, there exists $\theta_0=\theta^\ast$, known as the true parameter value. Let $\sigma_Z=\mathsf{Var}[S(Y,\theta_0^\top Z)|Z]$.

\begin{pro}\label{pro3}
Set $v=\mathsf{E}(\sigma_ZZ)$. Then the optimal $\pi$ to minimize the asymptotic variance of $v^\top\hat{\theta}$ is given by $\pi=c|S(Y,\theta_0^\top Z)|\wedge1$, where $c$ is a constant controlling the subsampling rate.
\end{pro}

\noindent Based on Proposition \ref{pro3}, given a pilot estimate $\tilde{\theta}$, we would propose the subsampling design $\pi_i=c|S(y_i,\tilde{\theta}^\top z_i)|\wedge1$, where $z_i=(1,x_i^\top)^\top$, $i=1,\ldots,n$. For the Logistic model with a binary response, we have $\psi(y,t)=p(t)^y(1-p(t))^{1-y}$ and $S(y,t)=(y-p(t))$. Then the proposed subsampling design becomes $\pi_i=c|y_i-p(\tilde{\theta}^\top z_i)|\wedge1$, which is exactly the LCC sampling proposed by Fithian and Hastie (2014). Thus, the LCC sampling can be viewed as a special case of the proposed surprise sampling design, with a somewhat narrow objective to achieve optimality in estimating a certain direction of the regression parameter vector. Surprise sampling, however, is better justified, since if the data analysis objective is altered, the sampling can be accordingly modified. We show in the next section that if Logistic model is correctly specified and the pilot estimate is consistent, the proposed HT type estimator $\hat{\theta}$ has the same asymptotic efficiency as the LCC estimator of Fithian and Hastie (2014).

Besides optimizing the prediction accuracy and estimation accuracy, other objectives can also be considered. The optimal selection probability $\pi$ adaptive to various objectives can be derived similarly to those in Proposition \ref{pro1} and \ref{pro2}. For instance, Wang et al. (2018) and Ai et al. (2018) mainly considered minimizing the conditional MSE of $\hat{\theta}$ given the full data. In our proposal, for minimizing the (unconditional) MSE, the optimal design is given by $\pi=(c\|A^{-1}g(D;\theta^\ast)\|)\wedge1$, which has the similar kernel part to the optimal form derived by Wang et al. (2018) under the Logistic model and Ai et al. (2018) under the generalized linear models.

\subsection{The algorithm}

Let $\tilde{\pi}_i$ be the kernel part of the proposed surprise sampling design, e.g. $\tilde{\pi}_i=\left\|\tilde{A}^{-1/2}g(d_i;\tilde{\theta})\right\|$ for prediction and $\tilde{\pi}_i=\left|v^\top\tilde{A}^{-1}g(d_i;\tilde{\theta})\right|$ for estimation. The implementation procedure of the surprise sampling is summarized in Algorithm \ref{alg_SS}.

\begin{algorithm}
\caption{Surprise Sampling}\label{alg_SS}
\begin{algorithmic}[1]
\REQUIRE data $d_i$, $i=1,\ldots,n$, a loss function $l(d;\theta)$, and a subsampling rate $r$.
\ENSURE the HT type estimate $\hat\theta$.
\STATE Get a pilot estimate $\tilde\theta$.
\STATE For $i=1,\ldots,n$, do:\\
 \ \ - \ evaluate $\tilde{\pi}_i$ and $\pi_i=c\tilde{\pi}_i\wedge1$, where $c$ is obtained by Algorithm \ref{alg_findC};\\
 \ \ - \ generate independent $\Delta_i\sim$ Bernoulli$(\pi_i)$.
\STATE Obtain a subsample $\{d_i: \Delta_i=1\}$.
\STATE Compute $\hat\theta$ by minimizing $\sum_{i=1}^n\Delta_il(d_i;\theta)/\pi_i$.
\end{algorithmic}
\end{algorithm}

The constant $c$ is obtained by Algorithm \ref{alg_findC}.

\begin{algorithm}
\caption{Find $c$ to reach the predetermined subsampling rate $r$}\label{alg_findC}
\begin{algorithmic}
\REQUIRE $\{\tilde{\pi}_i,i=1,\dots,n\}$ and the subsampling rate $r$.
\ENSURE the constant $c$ to control the subsample size.
\STATE Rearrange $\tilde{\pi}_i$'s in ascending order $\tilde{\pi}_{(1)}\leqslant\dots\leqslant\tilde{\pi}_{(n)}$, and compute $c_0=nr/\sum_{i=1}^n\tilde{\pi}_i$.
\IF{$c_0\tilde{\pi}_{(n)}\leqslant1$}
\STATE Set $c=c_0$.
\ELSE
\STATE Find $m$ such that $f(1/\tilde{\pi}_{(m+1)})\leqslant nr\leqslant f(1/\tilde{\pi}_{(m)})$ by the bisection method, where $f(x)=\sum_{i=1}^n(x\tilde{\pi}_i\wedge1)$.
\STATE Set $c=[nr-(n-m)]/\sum_{i=k+1}^m\tilde{\pi}_{(i)}$.
\ENDIF
\end{algorithmic}
\end{algorithm}

\section{Large sample properties}

In this section we discuss the asymptotic properties of the HT type estimator under the proposed surprise sampling designs. We first give out a general theorem on the large sample theory of $\hat{\theta}$ defined in (\ref{2.1}). The theorem shows that the consistency and the asymptotic normality of $\hat{\theta}$ hold when some conditions are assumed for the sampling design $\pi_i$ and the underlying distribution of the data.

\begin{thm}\label{thm1}
If conditions A1-A3 and C1-C4 listed in Appendix A.2 hold, then 1) $\hat{\theta}\overset{p}{\to}\theta^\ast$, where $\overset{p}{\to}$ means converging in probability (i.e., $\hat{\theta}$ is consistent) and 2) $\sqrt{n}(\hat{\theta}-\theta^\ast)\overset{d}{\to}N(0,A^{-1}V_\pi A^{-1})$, where $\overset{d}{\to}$ means converging in distribution and $N(\mu,\Sigma)$ stands for a normal vector with mean $\mu$ and variance-covariance matrix $\Sigma$.
\end{thm}

\noindent The key step to get the asymptotic distribution is to establish the expansion (\ref{3.0}). We give out the proof in Appendix A.2. It is worthwhile to mention that the consistency and the asymptotic normality of $\hat{\theta}$ does not require the pilot $\tilde{\theta}$ to be consistent nor to be independent of the full data, which, however, are both required in Fithian and Hastie (2014). The relaxations can be useful in practice, as we mentioned in Section 1. This is one of the main reasons that we propose to use the HT type estimator.

Based on Theorem \ref{thm1}, the asymptotic properties of the HT type estimator under the proposed surprise sampling designs can be derived immediately. For the surprise sampling, the specific forms of the design $\pi_i$ are given in (\ref{3.3}) or (\ref{3.4}), depending on the analysis purpose. We first consider the situation where the pilot $\tilde{\theta}$ is consistent. Here, the probability $\pi=(c\|A^{-1/2}g(D;\theta^\ast)\|)\wedge1$ for prediction and $\pi=(c|v^\top A^{-1}g(D;\theta^\ast)|\wedge1$ for estimation. The following corollary presents the asymptotic properties of $\hat{\theta}$ with the consistent pilot estimate.

\begin{cor}\label{cor1}
If conditions C1-C5 listed in Appendix A.2 hold and $\tilde{\theta}\overset{p}{\to}\theta^\ast$ (i.e., $\tilde{\theta}$ is consistent), then 1) $\hat{\theta}\overset{p}{\to}\theta^\ast$ (i.e., $\hat{\theta}$ is consistent) and 2) $\sqrt{n}(\hat{\theta}-\theta^\ast)\overset{d}{\to}N(0,A^{-1}V_\pi A^{-1})$.
\end{cor}

\noindent In Corollary \ref{cor1}, the probability $\pi$ in $V_\pi$ just takes the optimal form given in Proposition \ref{pro1} and \ref{pro2}. It means that when the pilot estimate is consistent, the proposed sampling designs given in (\ref{3.3}) or (\ref{3.4}) are optimal in the sense that the corresponding HT type estimator $\hat{\theta}$ asymptotically reaches the best prediction accuracy or estimation accuracy.

When the assumed model is correctly specified, the target parameter $\theta^\ast$ becomes the true parameter $\theta_0$. For the binary response  problem discussed in Fithian and Hastie (2014), we have shown that the LCC sampling is a special case of the proposed surprise sampling. Consider the sampling design $\pi_i=|y_i-p(\tilde{\theta}^\top z_i)|$, that is, the constant $c=1$. We have the following result for the corresponding $\hat{\theta}$.

\begin{cor}\label{cor2}
If the Logistic regression model $p(\theta^\top z)=\exp(\theta^\top z)/[1+\exp(\theta^\top z)]$ holds with $\theta_0$ being the true parameter value, $X$ does not concentrate on a hyperplane of dimension smaller than $q$, $\mathsf{E}(\|X\|^4/\pi)<\infty$, and $\tilde{\theta}\overset{p}{\to}\theta_0$, then $\sqrt{n}(\hat{\theta}-\theta_0)\overset{d}{\to}N(0,2\Sigma_{\mbox{\footnotesize full}})$, where $\Sigma_{\mbox{\footnotesize full}}$ stands for the asymptotic variance of the MLE for the full sample.
\end{cor}

\noindent Corollary \ref{cor2} means that under the correctly specified Logistic model, the proposed HT type estimator $\hat{\theta}$ is asymptotically as efficient as the LCC estimator of Fithian and Hastie (2014), as long as the pilot is consistent. Again, our result does not require the independence between the pilot and the full data. Both Corollary \ref{cor1} and \ref{cor2} are proved in Appendix A.2.

Then we consider the situation with an inconsistent pilot estimate. When the pilot estimate is not consistent, the LCC estimator is no longer consistent. However, based on Theorem \ref{thm1}, the HT type estimator can be still consistent and asymptotically normal as long as the pilot has a certain limit in probability. Specifically, suppose that $\tilde{\theta}$ converges in probability to a certain limit, denoted by $\bar{\theta}$. Let $\bar{A}=\mathsf{E}[G(D;\bar{\theta})]$ and $V_{\bar{\pi}}=\mathsf{E}[g(D;\theta^\ast)^{\otimes2}/\bar{\pi}]$, where $\bar{\pi}=(c\|\bar{A}^{-1/2}g(D;\bar{\theta})\|)\wedge1$ for prediction and $\bar{\pi}=(c|v^\top\bar{A}^{-1}g(D;\bar{\theta})|)\wedge1$ for estimation. For the inconsistent pilot, we have the following result.

\begin{cor}\label{cor3}
If conditions C1-C4 and C6 listed in Appendix A.2 hold and $\tilde{\theta}\overset{p}\to\bar{\theta}$, then 1) $\hat{\theta}\overset{p}{\to}\theta^\ast$ (i.e., $\hat{\theta}$ is consistent) and 2) $\sqrt{n}(\hat{\theta}-\theta^\ast)\overset{d}{\to}N(0,A^{-1}V_{\bar{\pi}} A^{-1})$.
\end{cor}

\noindent Again we give out the proof in Appendix A.2. When the pilot estimate is not consistent for $\theta^\ast$, the surprise sampling design is no longer optimal in the sense of minimizing the overall prediction error and the asymptotic variance. However, $\hat{\theta}$ is still consistent for $\theta^\ast$ and asymptotically normally distributed. Meanwhile, when $\bar\theta$ is not far away from $\theta^\ast$, which is a common case since the pilot is defined to be a ``good" guess of the target parameter, the surprise sampling design is still approximately optimal. Thus, the proposed estimation procedure is robust for the model mis-specification as well as the inconsistent pilot.

Finally, a plugged-in approach can be applied to estimate the asymptotic variance-covariance matrix of $\hat{\theta}$. Specifically, we define $\hat{A}=n^{-1}\sum_{i=1}^n\Delta_iG(d_i,\hat{\theta})/\pi_i$ and $\hat{V}_{\pi}=n^{-1}\sum_{i=1}^n\Delta_ig(d_i;\hat{\theta})^{\otimes2}/\pi_i^2$. A consistent estimator for the asymptotic variance-covariance matrix is given by $\hat{A}^{-1}\hat{V}_{\pi}\hat{A}^{-1}$.

\section{Simulation studies}

Here we conduct extensive simulation studies to examine the effectiveness of the proposed approach and make some comparison with the LCC sampling. The numerical studies mainly focus on various regression type problems, but as we have already mentioned, the whole procedure is general enough to be extended to unsupervised learning tasks. We first consider binary response where the proposed HT type estimator can be compared with the LCC estimator of Fithian and Hastie (2014).

\medskip

\noindent{\it Simulation 1: Correctly specified Logistic model}

In simulation 1 we consider the scenario where the Logistic model is correctly specified. We set $q=50$ and all the predictors $X$ are generated independently from the standard normal distribution. Given $X$, $Y$ is generated from a Bernoulli distribution with success probability $p(\theta^\top Z)$, where $\theta=(\alpha,\beta^\top)^\top$ with the first 25 components of $\beta$ being 1, the rest 25 being 0, and $\alpha$ being chosen to yield $\mathsf{P}(Y=1)\approx10\%$. The entire sample size $n$ is set to be $10^6$. For the pilot estimate, we use data drawn from the full sample with size $10^4$. We consider two types of consistent pilot estimates. The first one is to draw a random sample with uniform probabilities and get the Logistic MLE to be the pilot; the second one is to draw a 50-50 split case-control sample and apply the weighted case-control approach to get the pilot. For the subsampling design, we apply the LCC sampling $\pi_i=|y_i-p(\tilde{\theta}^\top z_i)|$, which is a special case of the proposed surprise sampling. Both the LCC estimator and the proposed HT type estimator based on the negative log-likelihood loss are obtained for comparison. The procedure is repeated for 1000 times. We record the squared bias and variance of the estimator for $\beta$, denoted by $\hat{\beta}$, over the 1000 realizations for each of the two methods under the two pilots, respectively. The results are summarized in Table 1.

%\begin{center}
%[Insert Table 1 here]
%\end{center}

{\begin{center}
Table 1. Comparison of LCC and HT estimate under the correctly specified Logistic model.

%\medskip

{\setlength{\tabcolsep}{1mm}
\begin{tabular}{lcc ccc ccc}\\ \hline\hline
&&Pilot     &&\multicolumn{2}{c}{Uniform MLE}     && \multicolumn{2}{c}{WCC}\\
&&Estimation&&$\widehat{\mbox{Bias}}^2 \ (\times10^{6})$&$\widehat{\mbox{Var}} \ (\times10^{3})$ &&$\widehat{\mbox{Bias}}^2 \ (\times10^{6})$&$\widehat{\mbox{Var}} \ (\times10^{3})$\\ \hline
&&LCC     && 3.888  & 3.682  && 3.347   & 3.613  \\
&&HT      && 4.052  & 3.767  && 3.477   & 3.676  \\
\hline\hline
\end{tabular}
{\footnotesize\begin{tablenotes}
\item[1]LCC: local case-control estimate; HT: Horvitz-Thompson type estimate; Uniform MLE: MLE with uniform sampling as pilot; WCC: weighted case-control estimate as pilot; $\widehat{\mbox{Bias}}^2$: $\widehat{\mbox{Bias}}^2=\|\mathsf{E}(\hat{\beta})-\beta_0\|^2$; $\widehat{\mbox{Var}}$: $\widehat{\mbox{Var}}=\sum_{j=1}^q\mathsf{Var}(\hat{\beta}_j)$.
\end{tablenotes}}}
\end{center}}

From Corollary \ref{cor2}, we know that when the Logistic model is correctly specified, the proposed HT estimator is asymptotically as efficient as the LCC estimator under the LCC sampling. From Table 1, we see that the behaviors of the two estimates are very close to each other, which coincides with the finding of the corollary. Also, the pilot method has little impact on the behavior of the proposed estimator, as long as it is consistent. 

It is worthwhile to mention that under the LCC sampling, the average size of the subsample is around 6.6\% of the entire sample size. Adding the sample size used for the pilot, it means that we use about 7.6\% of the full data size to reach half of the estimation efficiency. We also record the computation time. On the computer we conduct all the numerical studies (a laptop running macOS 10.14 with an Intel I7 processor and 16GB memory), it takes around 352 seconds to calculate a full sample MLE. The whole procedure of the subsample estimation, including obtaining the pilot, evaluating the sampling probabilities, drawing the subsample, and calculating the HT type estimate, however, takes around 19 seconds, only 5.4\% of the time for the full sample MLE.

\medskip

\noindent{\it Simulation 2: Incorrectly specified Logistic model with a consistent pilot}

In simulation 2 we turn to the scenario where the Logistic model is incorrectly specified. Here $q$ is set to be 5 and again all the predictors are generated independently from the standard normal distribution. Besides the linear combination of the 5 predictors, we add the quadratic term of the first predictor into the success probability of the Bernoulli distribution, but still fit the data by the usual Logistic model with linear terms only. Thus, the fitted model is incorrectly specified. The parameters are set to yield $\mathsf{P}(Y=1)\approx1\%$, so the data is more imbalanced than the last scenario. The entire sample size $n=10^6$, and the pilot sample size is $10^4$. We still use the Logistic MLE with uniform subsampling and the weighted case-control estimate as the pilot estimate, respectively. It is easy to see that the two pilots are consistent. Similar to simulation 1, the LCC sampling is applied and the LCC estimator and the HT estimator are obtained. The procedure is repeated for 1000 times and the results parallel to Table 1 are summarized in Table 2.

%\begin{center}
%[Insert Table 2 here]
%\end{center}

{\begin{center}
Table 2. Comparison of LCC and HT estimate under the incorrectly specified Logistic model with consistent pilot estimate.

%\medskip

{\setlength{\tabcolsep}{1mm}
\begin{tabular}{lcc ccc ccc}\\ \hline\hline
&&Pilot     &&\multicolumn{2}{c}{Uniform MLE}     && \multicolumn{2}{c}{WCC}\\
&&Estimation&&$\widehat{\mbox{Bias}}^2 \ (\times10^{6})$&$\widehat{\mbox{Var}} \ (\times10^{3})$ &&$\widehat{\mbox{Bias}}^2 \ (\times10^{6})$&$\widehat{\mbox{Var}} \ (\times10^{3})$\\ \hline
&&LCC     && 3.170  & 1.643  && 4.222   & 1.226  \\
&&HT      && 4.629  & 1.045  && 3.481   & 1.040  \\
\hline\hline
\end{tabular}
{\footnotesize\begin{tablenotes}
\item[1]LCC: local case-control estimate; HT: Horvitz-Thompson type estimate; Uniform MLE: MLE with uniform sampling as pilot; WCC: weighted case-control estimate as pilot; $\widehat{\mbox{Bias}}^2$: $\widehat{\mbox{Bias}}^2=\|\mathsf{E}(\hat{\beta})-\beta^\ast\|^2$; $\widehat{\mbox{Var}}$: $\widehat{\mbox{Var}}=\sum_{j=1}^q\mathsf{Var}(\hat{\beta}_j)$.
\end{tablenotes}}}
\end{center}}

From the results we see that the LCC estimate and the HT estimate are essentially unbiased for the target parameter $\beta^\ast$. Meanwhile, in this case, the empirical variance of the proposed HT estimate is smaller than that of the LCC estimate. Under the LCC sampling, the average size of the subsample is around 1.9\% of the entire sample size.

\medskip

\noindent{\it Simulation 3: Incorrectly specified Logistic model with an inconsistent pilot}

In simulation 3 we compare the two estimators under incorrectly specified Logistic with an inconsistent pilot estimate. The data generation scheme is the same as that in simulation 2. For the pilot estimate, we still draw a subsample of size $10^4$ with uniform probabilities. However, the difference here is that we use the probit MLE to be the pilot estimate. Then the resulting pilot estimate is no longer consistent for the target parameter. Using this inconsistent pilot, the LCC sampling is applied and the LCC estimator and the HT estimator are obtained. The procedure is repeated for 1000 times. Here, besides the results of the two estimates, we also report the squared bias and variance of the pilot estimate. The results are summarized in Table 3.

%\begin{center}
%[Insert Table 3 here]
%\end{center}

{\begin{center}
Table 3. Comparison of LCC and HT estimate under the incorrectly specified Logistic model with inconsistent pilot estimate.

%\medskip

{\setlength{\tabcolsep}{1mm}
\begin{tabular}{lcc ccc ccc}\\ \hline\hline
&&Estimation&&$\widehat{\mbox{Bias}}^2 \ (\times10^{4})$&&$\widehat{\mbox{Var}} \ (\times10^{4})$\\ \hline
&&Pilot$^{\mbox{\scriptsize p}}$ && 935.5 && 102.3 \\
&&LCC                && 6.432  && 7.138\\
&&HT                 && 0.042  && 7.273\\
\hline\hline
\end{tabular}
{\footnotesize\begin{tablenotes}
\item[1]Pilot$^{\mbox{\scriptsize p}}$: pilot estimate with the probit fitting; LCC: local case-control estimate; HT: Horvitz-Thompson type estimate; $\widehat{\mbox{Bias}}^2$: $\widehat{\mbox{Bias}}^2=\|\mathsf{E}(\hat{\beta})-\beta^\ast\|^2$; $\widehat{\mbox{Var}}$: $\widehat{\mbox{Var}}=\sum_{j=1}^q\mathsf{Var}(\hat{\beta}_j)$.
\end{tablenotes}}}
\end{center}}

The consistency of the LCC estimator relies on the consistency of the pilot, but the that of the proposed HT estimator does not, as Corollary \ref{cor3} illustrates. From Table 3, we see that the pilot estimate is clearly biased. Compared with the HT estimator, the bias of the LCC estimator is quite obvious. Although the squared bias of the LCC estimator is not large in the absolute value, it is of the same order with its variance. The proposed HT estimator, however, has much smaller squared bias relative to its variance, showing positive evidence for Corollary \ref{cor3}. This is an advantage of the HT estimator over the LCC one.

\medskip

\noindent{\it Simulation 4: Comparison of LCC sampling and surprise sampling}

Proposition \ref{pro3} shows that the LCC sampling is a special case of the proposed surprise sampling design, and is optimal when the target is to estimate $v^\top\theta^\ast$ with $v=\mathsf{E}(\sigma_ZZ)$. However, when the target direction changes, the LCC sampling is not adaptively optimal, but our surprise sampling design is. In this simulation, we again generate data by the same scheme as that in simulation 2, but suppose that the main concern is to estimate the regression coefficient of the first predictor, i.e., $\beta_1^\ast$. Equivalently speaking, the target direction $v$ here is a 6-dimensional column vector with the second element being 1 and all the others being 0. For this target, the LCC sampling is not optimal. Following Proposition \ref{pro2}, the optimal design is given by $\pi_i=(c|v^\top\tilde{A}^{-1}z_i(y_i-p(\tilde{\theta}^\top z_i))|)\wedge1$, where $\tilde{A}=n^{-1}\sum_{i=1}^np(\tilde{\theta}^\top z_i)(1-p(\tilde{\theta}^\top z_i))z_iz_i^\top$. The two subsampling designs are compared, with $c$ decided to yield comparable average subsample sizes between the two designs. We still use the same two pilot estimates as simulation 2 and consider three estimation approaches: the LCC estimator, the HT estimator under the LCC sampling, and the HT estimator under the optimal surprise sampling. The procedure is repeated for 1000 times. For the HT estimator, besides the squared bias and variance, we also report the average of the variance estimates and the empirical coverage percentage of the 95\% Wald confidence interval. The results are summarized in Table 4.

%\begin{center}
%[Insert Table 4 here]
%\end{center}

{\begin{center}
Table 4. Estimation of $\beta_1^\ast$ by using local case-control sampling and surprise sampling under the Logistic model.

%\medskip

{\setlength{\tabcolsep}{1mm}
\begin{tabular}{lcc ccc ccc ccc ccc ccc}\\ \hline\hline
&&Pilot     &&\multicolumn{4}{c}{Uniform MLE}     && \multicolumn{4}{c}{WCC}\\
&&          &&Bias$^2$ &Var &Var Est. & CP &&Bias$^2$ &Var &Var Est. & CP \\
&&Estimation&&\small{$(\times10^{6})$}&\small{$(\times10^{4})$}&\small{$(\times10^{4})$}&\small{(\%)}  &&\small{$(\times10^{6})$}&\small{$(\times10^{4})$}&\small{$(\times10^{4})$}&\small{(\%)} \\ \hline
&&LCC        && 0.473   & 7.799  &  -    & -    && 1.295 & 3.417 &  -    & -    \\
&&HT-LCC     && 1.669   & 1.684  & 1.444 & 94.1 && 0.139 & 1.540 & 1.394 & 94.4\\
&&HT-optimal && 0.577   & 1.134  & 0.978 & 92.0 && 0.065 & 1.129 & 0.943 & 92.5 \\
\hline\hline
\end{tabular}
{\footnotesize\begin{tablenotes}
\item[1]LCC: local case-control estimate; HT-LCC: Horvitz-Thompson type estimate under the local case-control sampling; HT-optimal: Horvitz-Thompson type estimate under the optimal surprise sampling; Bias$^2$: average of the squared bias; Var: empirical variance; Var Est.: average of the variance estimate; CP: coverage probability of the 95\% Wald confidence interval; Uniform MLE: MLE with uniform sampling as pilot; WCC: weighted case-control estimate as pilot.
\end{tablenotes}}}
\end{center}}

The variance of the HT estimator under the optimal surprise sampling is much smaller than that of the LCC estimator. By using comparable subsample sizes (around 1.9\% of the entire sample size), the relative efficiency of the HT estimator under the optimal sampling to the LCC estimator is at least 3 in our situation. Also, the HT estimator under the optimal sampling is more efficient than that under the LCC sampling, which is expected according to Proposition \ref{pro2}. The plugged-in variance estimate and the Wald confidence interval give out reasonable performance.

\medskip

In the remaining studies we turn to more general regression models with counting and continuous responses, where the LCC approach is no longer available.

\medskip

\noindent{\it Simulation 5: Counting response under log-linear model}

In simulation 5 we consider the counting response. We set $q=2$ and the two predictors $X=(X_1,X_2)$ are generated independently from the standard normal distribution. Two schemes for generating $Y$ given $X$ are considered. In the first one, $Y$ is generated from a Poisson distribution with mean $\exp(\alpha+\beta_1X_1+\beta_2X_2)$, while in the second one, $Y$ is generated from a Poisson distribution with mean $\exp(\alpha+\beta_1X_1+\beta_2X_2+\beta_3X_1^2)$. Thus, the Poisson log-linear model only with the linear terms is correctly specified for the first scheme, but misspecified for the second. In both schemes, the parameters are set to yield $\mathsf{P}(Y\leqslant1)\approx93\%$, implying that the generated data is imbalanced in the sense that most responses are 0 or 1. The entire sample size $n=10^5$. We draw a pilot sample of size $1000$ from the entire sample with uniform probabilities and fit the Poisson log-linear MLE to be the pilot, denoted by $\tilde{\theta}=(\tilde{\alpha},\tilde{\beta}_1,\tilde{\beta}_2)^\top$. According to Proposition \ref{pro3}, we use the surprise sampling design $\pi_i=(c|y_i-\exp(\tilde{\theta}^\top z_i)|)\wedge1$, where $z_i=(1,x_{1i},x_{2i})^\top$ and $c$ is set to make the subsample size equal to $1000$. The proposed HT type estimator based on the negative log-likelihood loss is calculated. For comparison, we draw a uniform subsample of size $2000$ to calculate the Poisson log-linear MLE. The size is $2000$ since the proposed surprise sampling needs to pay for its pilot sample. The procedure is repeated for 1000 times. The squared bias and variance of the two estimators are recorded. For the HT estimator, we also record the average of the variance estimates and the empirical coverage percentage of the 95\% Wald confidence interval. The results are summarized in Table 5.

%\begin{center}
%[Insert Table 5 here]
%\end{center}

{\begin{center}
Table 5. Estimation results under the Log-linear model.

%\medskip

{\setlength{\tabcolsep}{1mm}
\begin{tabular}{lcc ccc ccc ccc ccc ccc}\\ \hline\hline
&&              &&Estimation&&\multicolumn{2}{c}{Sub-MLE}  && \multicolumn{4}{c}{HT}\\
&&Model         &&          &&Bias$^2$ &Var &&Bias$^2$ &Var &Var Est. & CP \\
&&specification &&Parameter &&\small{$(\times10^{6})$}&\small{$(\times10^{3})$} &&\small{$(\times10^{6})$}&\small{$(\times10^{3})$}&\small{$(\times10^{3})$}&\small{(\%)} \\ \hline
&&              &&$\alpha$   && 94.49 & 8.414 && 3.145 & 4.334 & 4.115 & 94.4    \\
&&Correct       &&$\beta_1$  && 5.979 & 3.200 && 0.307 & 1.034 & 0.963 & 94.6    \\
&&              &&$\beta_2$  && 0.822 & 3.083 && 2.292 & 0.958 & 0.954 & 95.5    \\
&&              &&$\alpha$   && 45.02 & 11.95 && 12.09 & 5.041 & 4.976 & 94.2    \\
&&Incorrect     &&$\beta_1$  && 20.79 & 4.437 && 0.267 & 1.437 & 1.528 & 95.7    \\
&&              &&$\beta_2$  && 4.642 & 6.372 && 8.191 & 1.714 & 1.774 & 95.4    \\ \hline\hline
\end{tabular}
{\footnotesize\begin{tablenotes}
\item[1]Full MLE: full sample MLE; Sub-MLE: MLE with uniform subsample of size 2000; HT: Horvitz-Thompson type estimate under the surprise sampling; Bias$^2$: average of the squared bias; Var: empirical variance; Var Est.: average of the variance estimate; CP: coverage probability of the 95\% Wald confidence interval.
\end{tablenotes}}}
\end{center}}

From the results, we find out that no matter whether the model is correctly specified or not, the HT estimator based on the proposed surprise sampling is essentially unbiased for the target parameter, and is much more efficient than the uniform subsample MLE with a comparable sample size. The plugged-in variance estimate and the Wald confidence interval give out satisfactory performances.

\medskip

\noindent{\it Simulation 6: Continuous response under linear model}

In the last simulation we consider the continuous response. Still $q$ is set to be 2, and the two predictors $X=(X_1,X_2)$ are generated independently from a normal distribution with zero mean and variance 0.01. Two schemes for generating $Y$ given $X$ are considered. In the first one, $Y$ is generated from a normal distribution with mean $\alpha+\beta_1X_1+\beta_2X_2$ and variance 0.01, while in the second one, $Y$ is generated from Poisson distribution with mean $\alpha+\beta_1X_1+\beta_2X_2+\beta_3X_1^2$ and variance 0.01. Thus, the linear regression model only with the linear terms is correctly specified for the first scheme, but misspecified for the second. In both schemes, the parameters are set to yield $\mathsf{P}(-0.5<Y<0.5)\approx99.6\%$, implying that most responses are near zero. The entire sample size $n=10^5$. We draw a uniform pilot sample of size $1000$ from the entire sample and fit the normal linear regression MLE to be the pilot $\tilde{\theta}$. For the surprise sampling, we use $\pi_i=(c|y_i-\tilde{\theta}^\top z_i|)\wedge1$, where $c$ is set to make the subsample size equal to $1000$. The proposed HT type estimator based on the least squared loss is calculated. Similar to the last simulation, the MLE based on a uniform subsample of size 2000 is also obtained for comparison. The parallel results are summarized in Table 6.

%\begin{center}
%[Insert Table 6 here]
%\end{center}

{\begin{center}
Table 6. Estimation results under the linear model.

%\medskip

{\setlength{\tabcolsep}{1mm}
\begin{tabular}{lcc ccc ccc ccc ccc ccc}\\ \hline\hline
&&              &&Estimation&&\multicolumn{2}{c}{Sub-MLE}  && \multicolumn{4}{c}{HT}\\
&&Model         &&          &&Bias$^2$ &Var &&Bias$^2$ &Var &Var Est. & CP \\
&&specification &&Parameter &&\small{$(\times10^{8})$}&\small{$(\times10^{4})$} &&\small{$(\times10^{8})$}&\small{$(\times10^{4})$}&\small{$(\times10^{4})$}&\small{(\%)} \\ \hline
&&              &&$\alpha$   && 2.178 & 0.103 && 1.482 & 0.067 & 0.066 & 94.5 \\
&&Correct       &&$\beta_1$  && 0.774 & 10.77 && 8.392 & 6.651 & 6.665 & 95.5 \\
&&              &&$\beta_2$  && 66.70 & 10.05 && 24.05 & 6.667 & 6.712 & 95.9 \\
&&              &&$\alpha$   && 2.459 & 0.104 && 2.016 & 0.067 & 0.066 & 94.6 \\
&&Incorrect     &&$\beta_1$  && 1.890 & 10.82 && 53.54 & 6.881 & 6.779 & 94.8 \\
&&              &&$\beta_2$  && 5.766 & 10.14 && 2.796 & 6.745 & 6.818 & 95.9 \\ \hline\hline
\end{tabular}
{\footnotesize\begin{tablenotes}
\item[1]Sub-MLE: MLE with uniform subsample of size 2000; HT: Horvitz-Thompson type estimate under the surprise sampling; Bias$^2$: average of the squared bias; Var: empirical variance; Var Est.: average of the variance estimate; CP: coverage probability of the 95\% Wald confidence interval.
\end{tablenotes}}}
\end{center}}

The observations of the results are basically similar to those of Table 5, so we omit repeating the details. The last two simulations show that the proposed surprise sampling and the corresponding HT estimator are quite effective for counting and continuous data.

\section{Applications}

\noindent{\it Web Spam data}

We first apply the proposed approach to the Web Spam data available on the LIBSVM website and originally from Webb, Caverlee, and Pu (2006). The same data was also analyzed by Fithian and Hastie (2014) using the LCC approach for spam filtering. The data consists of 350000 web pages with about 60\% are labeled as ``web spam" that designed to manipulate search engines rather than display legitimate content. Following Fithian and Hastie (2014), we use frequency of the 99 unigrams that appeared in at least 200 documents as features, log-transformed with an offset to reduce skew. This data set is marginally balanced, but has considerable conditional imbalance in the sense that for some feature values, the web label is easy to predict.

We keep the same subsampling procedure as that in Fithian and Hastie (2014). The LCC sampling design is used and the corresponding selection percentage is about 10\% for this data. To assess the sampling distribution of the estimators, we repeatedly take uniform subsample of size $n=100000$ from the full data for 100 times. In each replication, we use the weighted case-control approach adopted in {\it Simulation 1} with size 10000 to get the pilot estimate. By the LCC design, the subsample size for parameter estimation is also around 10000. We fit the full sample MLE (of size $100000$), the LCC estimate, and the proposed HT estimate. Following Corollary \ref{cor2}, under the Logisitic model, the asymptotic variance of the proposed HT estimate is twice the variance of the full sample MLE, and is the same as that of the LCC estimate. Since in this real data, it is very likely the Logistic model is incorrectly specified, as Fithian and Hastie (2014) mentioned, the asymptotic variance of the HT estimate and the LCC estimate should be a little bit more than two times the variance of the MLE. On the other hand, the subsample size is around 20000 (including the sample for pilot estimate), i.e., 20\% of the full sample size. Then a uniform subsample of size 20000 should yield variance roughly 5 times that of the full sample.

%\begin{center}
%[Insert Figure 1 here]
%\end{center}

\begin{center}
\includegraphics[scale=0.55]{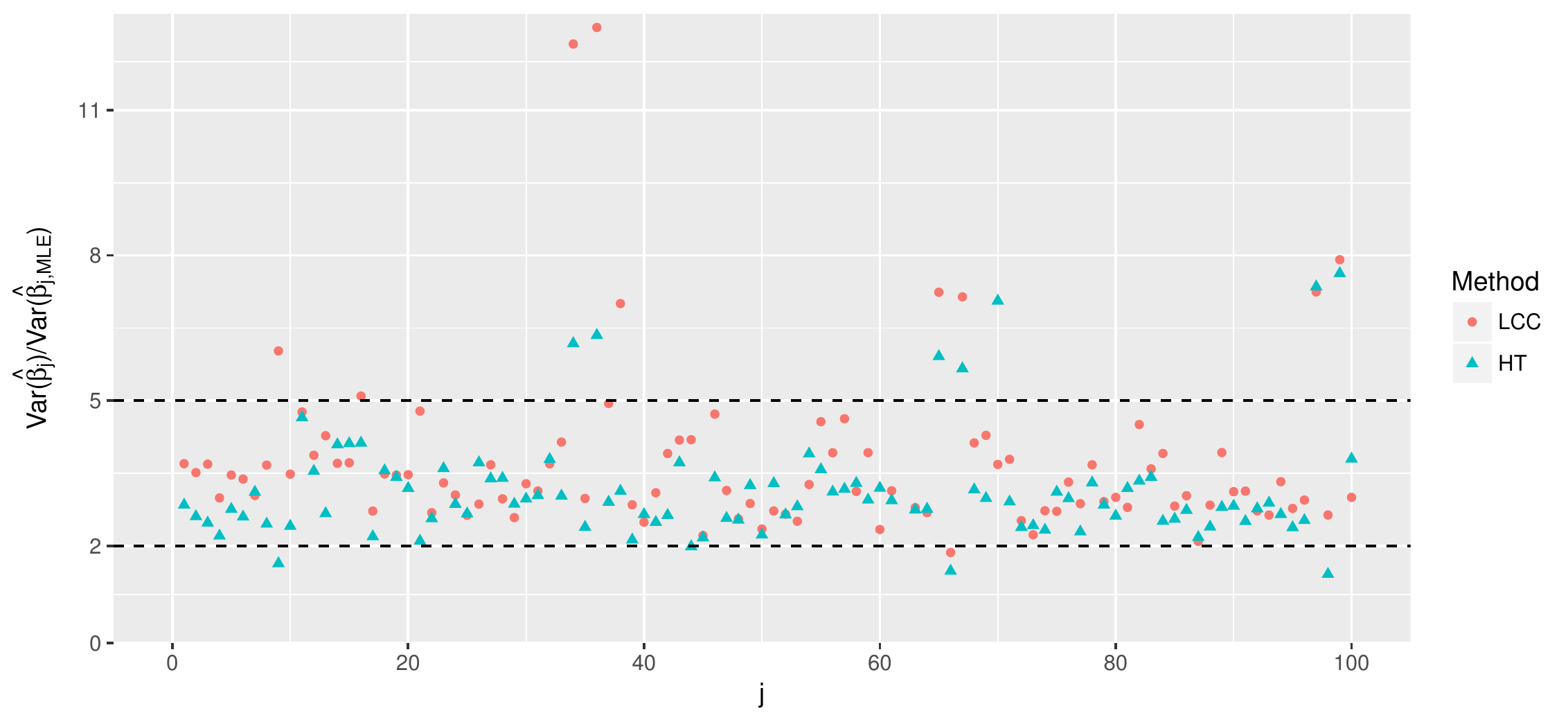}

Figure 1. Relative variance of the subsampling estimate ($\hat{\beta}_{j}$) to the full sample MLE ($\hat{\beta}_{j,{\footnotesize \mbox{MLE}}}$). The triangle is for the HT type estimate. The round is for the LCC estimate.
\end{center}

Figure 1 shows the relative variances of the subsampling estimates to the full sample MLE. Specifically, the horizontal axis indexes the 100 regression coefficients to fit, and the vertical axis stands for the variance of each estimated coefficient relative to that of the full-sample MLE of the same coefficient. The triangle is for the HT estimate and the circle is for the LCC estimate. We find that most relative variances of the two subsampling estimates are slightly larger than 2, as expected, but are substantially smaller than 5, implying the LCC sampling is more efficient than the uniform subsampling with a comparable sample size. Moreover, for most estimated coefficients, the relative variance of the HT estimate is smaller than that of the LCC estimate, implying that the proposed HT estimator is slightly more efficient than the LCC estimator. This is consistent with the finding of {\it Simulation 2}. The average computation time of 100 replications for the full sample MLE is 66.64 seconds. The average computation time for the LCC estimate (including the subsampling design implementation) is 19.32 seconds and for the HT estimate is 20.04 seconds.

\medskip

\noindent{\it Micro-blog data}

In the second example we use the proposed method to analyze a data set of micro-blog, a Chinese version of twitter. We collect 625895 tweets posted on Sina micro-blog during January to March, 2018. For each tweet, the time of posting (in hour), the number of comments, retweets, and likes are recorded. Moreover, we also record the gender, the number of followers, fans, and the number of tweets posted before of the blogger who posted the tweet. We set the number of likes to be the response and the other variables to be the predictors. Since the number of likes is a counting variable, the log-linear model is used to do analysis. The response, ranging from 0 to tens of thousands, is very right skewed, i.e., imbalanced. The mean of response is 1038 while the median is 72. Around 55\% of the tweets have the number of likes less than 100, while 2.11\% of the tweets have the number of likes larger than 10000. In Figure 2, we show the histogram of the response.

%\begin{center}
%[Insert Figure 2 here]
%\end{center}

\begin{center}
\includegraphics[scale=0.5]{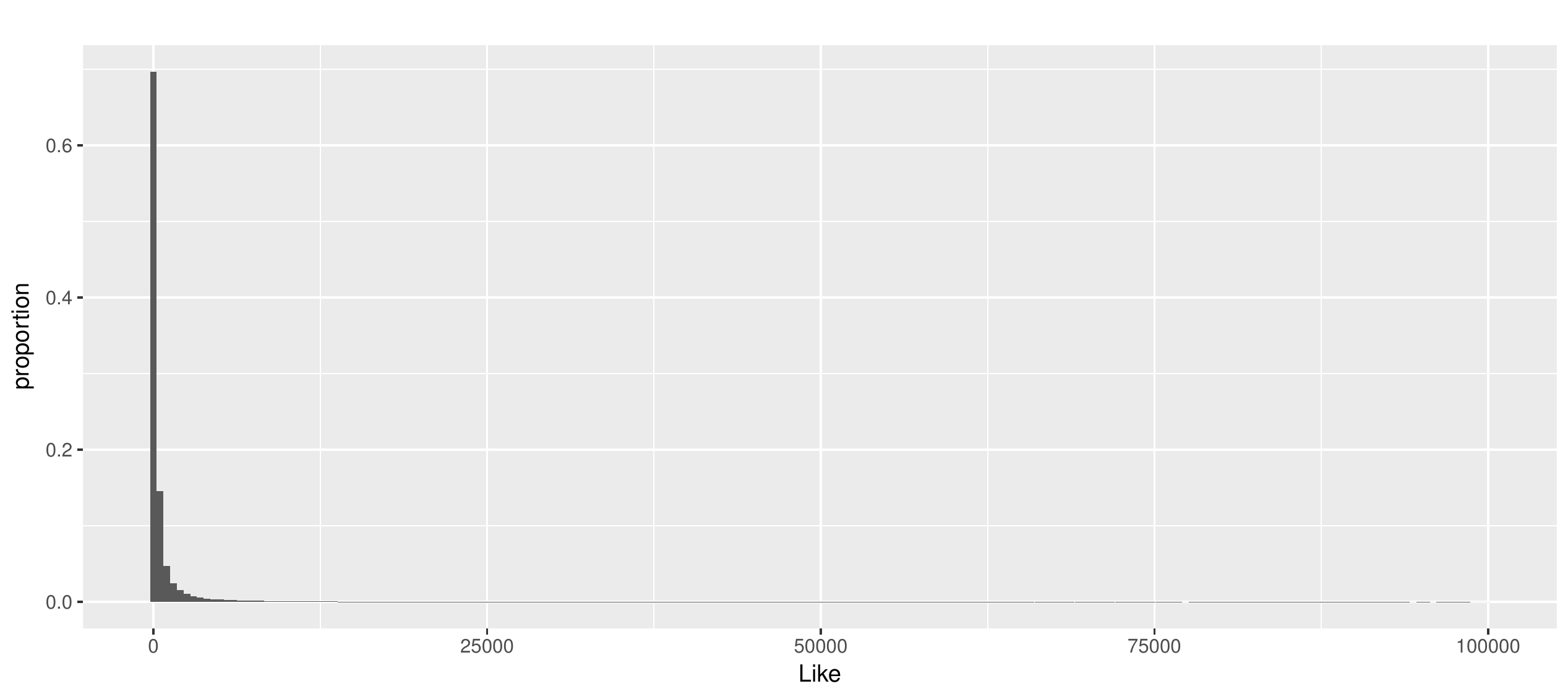}

Figure 2. Histogram of the number of likes in Micro-blog data.
\end{center}

In this analysis we focus on predicting the response by the predictors. Usually, the full data should be randomly separated into the training part for training model and the testing part for measuring prediction accuracy. To increase the stability of the results, here we use a ten-fold cross-validation style approach, that is, the full data is randomly split into ten parts and in each time, nine parts of the data serve as the training set and the rest one part serves as the testing set. Each time we train the log-linear model based on the training set and then calculate the root mean square error (RMSE) of the predicted values on the testing set, i.e., RMSE$=\sqrt{n_t^{-1}\sum_{i\in I_t}(y_i-\hat{y}_i)^2}$, where $I_t$ is the indicator set of the testing data, $n_t$ is the size of $I_t$, $y_i$ is the response value, and $\hat{y}_i$ is the predicted value. Finally, the average of the ten RMSEs, called ARMSE, is treated as the measure of the prediction accuracy. In the training set, we use the same sampling design as that in {\it Simulation 5} to get the subsample, i.e., $\pi_i=(c|y_i-\exp(\tilde{\theta}^\top z_i)|)\wedge1$, where $y_i$ is the response, $x_i$ is the standardized predictor vector, and $\tilde{\theta}$ is the pilot estimate which is the MLE fitted with a uniform subsample of size 10000. By taking a proper value of $c$, the size of the surprise subsample is set to be 10000. To make comparison, we also draw a uniform subsample of size 20000, which has a comparable sample size to the surprise subsample, to train the model by MLE. The prediction result of the full data MLE serves as the benchmark.

The ARMSE based on the full data is $6.848\times10^3$. The ARMSE based on surprise sampling data is $7.194\times10^3$, which is quite close to that of the full data. The ARMSE based on the uniform subsample data is, however, $2.367\times10^{13}$, which is much greater than that of the surprise sampling. One of the main reasons of such result is the right-skewness of the response data. With the subsampling percentage used here, the uniform subsampling has little chance to pick out the tweets with the extremely large responses, say, larger than 10000. Because of the significant influence of the large responses on the model fitting, the training models based on the uniform subsample data in some folds have quite different coefficients for some predictors from the full data models. Consequently, the prediction based on the uniform subsample models are very inaccurate for some responses in the testing set, resulting in extremely large RMSEs. By contrast, the surprise sampling is more likely to select those large responses, or in other words, the ``surprising" responses, into the subsample due to its design motivation (the selection probability can be 1 for those extremely large responses compared with the pilot prediction). Consequently, the training models based on surprise sampling data take the large responses into consideration and eventually they have similar coefficients estimation and prediction performance with the full data models. The average computation time of 10 folds for the full data is 97.45 seconds. For the surprise sample data, the average computation time is 4.22 seconds and for the uniform sample data, is 3.54 seconds.

\section{Conclusion and discussion}

We develop an improved version of the LCC sampling, called surprise sampling design, to reduce computational burden and retain good prediction or estimation performances. The proposed sampling design is flexibly adaptive to different objectives. For a specific objective, the design has the corresponding optimality. The LCC sampling design, in this sense, can be viewed as a special case of the proposed surprise sampling design. For parameter estimation, we propose the HT type estimation approach. The resulting estimator is consistent and asymptotically normally distributed, without requiring the pilot estimator to be consistent or to be independent of the full data. For the binary response, if the Logistic model is correctly specified and the pilot is consistent, the HT estimator is asymptotically as efficient as the estimator of Fithian and Hastie (2014) under the LCC sampling. The surprise sampling design and the HT estimation can be extended to more general responses such as counting and continuous response.

The proposed subsampling approach is more of a working principle for efficient data analysis which can be applied to various statistical learning problems. In the numerical studies we mainly focus on regression. However, as we mention many times, the optimal sampling design, the algorithm described in Section 3, and the theories discussed in Section 4 can be readily extended to unsupervised learning tasks. The main reason is that our approach setup essentially only requires a specific loss function and a finite-dimensional parameter. One example of unsupervised learning is briefly given in Section 2. Some more detailed discussion on the approach and more meaningful applications of unsupervised learning are of great interest.  

There are several other directions worth further research. Firstly, in our proposed approach, the pilot estimate is a guess of the target parameter, so they are of the same dimension. In some real application, people may just want to use a relatively simpler model and smaller sample size to get the pilot quickly and then apply a more complicated model to estimate the target parameter or do prediction. Then the pilot model used is different from the fitted model and the pilot is no longer a formal guess of the target parameter. Thus, the surprise sampling design needs adjustment and the optimality of the design requires more investigation. Secondly, the HT type estimator defined in (\ref{2.1}) is not semiparametrically efficient, but an immediate variant of the HT estimator is. The surprise sampling can be defined accordingly with the variant, in which the sampling probability $\pi_i$ is estimated by the kernel smoothed approach. It is possible to further improve the estimation efficiency. Thirdly, when the dimension of the target parameter is high, one may introduce regularization penalties, such as lasso and ridge, into the objective function in (\ref{2.1}). Based on the proposed framework, the regularization can be easily incorporated. The large sample properties of the resulting estimator needs more exploration.

\appendix

\section{Appendix}

In Appendix, we prove the propositions and theorems in Section 3 and 4.

\subsection{Proof of the propositions}

\noindent{\it Proof of Proposition \ref{pro1}:} \ Set $\tilde{\pi}=\|A^{-1/2}g(D;\theta^\ast)\|$. By the Cauchy-Schwarz inequality,
\begin{eqnarray*}
\mathsf{E}(\tilde{\pi})=\mathsf{E}\left(\sqrt{\pi}\cdot\frac{\tilde{\pi}}{\sqrt{\pi}}\right)\leqslant \sqrt{\mathsf{E}(\pi)\cdot\mathsf{E}\left(\frac{\tilde{\pi}^2}{\pi}\right)},
\end{eqnarray*}
where the equality holds if and only if $\sqrt{\pi}\propto \tilde{\pi}/\sqrt{\pi}$ which is equivalent to $\pi=c\cdot\tilde{\pi}$ with $c$ being a constant. From this inequality we get an achievable lower bound of (\ref{3.2}), that is,
\begin{eqnarray*}
\min_{\pi(\tilde{\pi})}\mathsf{E}\left(\frac{\tilde{\pi}^2}{\pi}\right)=\left.\left[\left(\mathsf{E}(\tilde{\pi})\right)^2\cdot\frac{1}{\mathsf{E}(\pi)}\right]\right|_{\pi=c\tilde{\pi}}=\frac{\mathsf{E}(\tilde{\pi})}{c}.
\end{eqnarray*}
By the Karush-Kuhn-Tucker (KKT) conditions, with the constraint $0\leqslant\pi\leqslant1$, the optimal $\pi$ must be given by $c\tilde{\pi}\wedge1$. Then with the constraint $\mathsf{E}(\pi)\leqslant r$, $c$ is determined as the largest constant such that $\mathsf{E}(c\tilde{\pi}\wedge1)\leqslant r$. \qed

\medskip

\noindent{\it Proof of Proposition \ref{pro2}:} \ Set $\tilde{\pi}=|v^\top A^{-1/2}g(D;\theta^\ast)|$. Note that the asymptotic variance of $\sqrt{n}v^\top(\hat{\theta}-\theta^\ast)$ which we want to minimize is $v^\top A^{-1}V_\pi A^{-1}v = \mathsf{E}(v^\top A^{-1}g(D;\theta^\ast)^{\otimes2}A^{-1}v^\top/\pi) = \mathsf{E}(|v^\top A^{-1}g(D;\theta^\ast)|^2/\pi)$. Then the proof is exactly the same with that of Proposition 1. \qed

\medskip

\noindent{\it Proof of Proposition \ref{pro3}:} \  Let $e_1$ be a $p$-dimensional unit vector whose first component is one and all the others are zero. We first prove that
\begin{eqnarray}\label{eq_base}
v^\top A^{-1}=e_1^\top.
\end{eqnarray}	
When $\theta=\theta_0$ we have that
\begin{eqnarray*}
\mathsf{E}\left(G(Y,X;\theta)|X\right)&=& -\mathsf{E}\left(\left.\frac{\partial^2}{\partial\theta\partial\theta^\top}\log\psi(Y,\theta^\top Z)\right|Z\right)\\
&=&\mathsf{Cov}\left(\left.\frac{\partial}{\partial\theta}\log\psi(Y,\theta^\top Z)\right|Z\right)=\mathsf{Cov}\left(S(Y,\theta^\top Z)Z\mid Z\right)\\
&=&\mathsf{E}\left[\left(S(Y,\theta^\top Z)Z-\mathsf{E}\left(S(Y,\theta^\top Z)Z\mid Z\right)\right)^{\otimes2}\mid Z\right] \\
&=&\mathsf{E}\left[\left(S(Y,\theta^\top Z)-\mathsf{E}\left(S(Y,\theta^\top Z)\mid Z\right)\right)^2|Z\right]ZZ^\top\\
&=&\mathsf{Var}\left[S(Y,\theta^\top Z)|Z\right]ZZ^\top,
\end{eqnarray*}
where $\mathsf{Cov}$ stands for the variance-covariance matrix of a random vector. Hence, $A=\mathsf{E}[G(Y,X;\theta_0)]=\mathsf{E}\left[\mathsf{Var}\left(S(Y,\theta_0^\top Z)\mid Z\right)ZZ^\top\right]=\mathsf{E}\left(\sigma_Z ZZ^\top\right)$.
Since the first component of $Z$ is constant 1, we have that $e_1^\top Z=1$. Also, note that $\sigma_Z$ is a scalar. Thus, $e_1^\top A=\mathsf{E}\left(\sigma_Z e_1^\top ZZ^\top\right)=\mathsf{E}\left(\sigma_Z Z^\top\right)=v^\top$, which leads to (\ref{eq_base}).

Next, we have that
\begin{eqnarray*}
v^\top A^{-1}g(y,x;\theta_0)&=&e_1^\top (-S(y,\theta_0^\top z)z)=-e_1^\top S(y,\theta_0^\top z)(1,x^\top)^\top \\
&=&-e_1^\top (S(y,\theta_0^\top z),S(y,\theta_0^\top z)x^\top)^\top = -S(y,\theta_0^\top z).
\end{eqnarray*}
On one hand, by Proposition \ref{pro2} we know that the asymptotic variance of $v^\top\hat{\theta}$ based on the optimal $\pi$ is
\begin{equation}\label{eq_optvar}
\mathsf{E}\left(\left(\frac{1}{c}\left|v^\top A^{-1}g(Y,X;\theta^*)\right|\right) \vee \left(v^\top A^{-1}g(Y,X;\theta^*)\right)^2\right),
\end{equation}
which equals to $\mathsf{E}\left(\left(|S(Y,\theta_0^\top Z)|/c\right) \vee \left(S^2(Y,\theta_0^\top Z)\right)\right)$ because of (\ref{eq_base}). On the other hand, the asymptotic variance of $v^\top\hat{\theta}$ based on the $\pi$ defined in the proposition is
\begin{eqnarray*}
\left.\mathsf{E}\left(\frac{1}{\pi}\left|v^\top A^{-1}g(Y,X;\theta^*)\right|^2\right)\right|_{\pi=c|S(Y,\theta_0^\top Z)|\wedge1} = \mathsf{E}\left(\left(\frac{1}{c}\left|S(Y,\theta_0^\top Z)\right|\right) \vee \left(S^2(Y,\theta_0^\top Z)\right)\right)
\end{eqnarray*}
which is equal to (\ref{eq_optvar}). Therefore, $\pi(Y,X;\theta_0)=c|S(Y,\theta_0^\top Z)|\wedge1$ is the optimal $\pi$ to minimize the asymptotic variance of $v^\top\hat{\theta}$. \qed

\subsection{Proof of the theorem and corollaries}

In order to make the presentation more concise, we use some abbreviation of the notation. Specifically, for $i=1,\ldots,n$, let $l_i(\theta)=l(d_i;\theta)$, $g_i(\theta)=g(d_i;\theta)$, and $G_i(\theta)=G(d_i;\theta)$. Let $R_n(\theta)=n^{-1}\sum_{i=1}^n\Delta_il_i(\theta)/\pi_i$ and denote the $\sigma$-algebra generated by the observed data and the pilot estimate $\tilde{\theta}$ by $\mathcal{F}_n$.

Some conditions are needed. The first sets of conditions are mainly about the subsampling probabilities $\pi_i$'s.

\noindent{\it A1}. \ $n^{-1}\sum_{i=1}^n(\sup_{\theta\in\Theta}l_i(\theta))^2/\pi_i=O_p(1)$, where $\Theta\in\mathbb{R}^p$ is the parameter space.

\noindent{\it A2}. \ $n^{-1}\sum_{i=1}^n\|G_i(\theta^\ast)\|^2/\pi_i=O_p(1)$.

\noindent{\it A3}. \ There exists a probability $\pi$ that may depend on $D$ such that $\mathsf{E}(\|g(D;\theta^\ast)\|^2/\pi)<\infty$ and $n^{-1}\sum_{i=1}^n(1/\pi_i-1)g_i(\theta^\ast)^{\otimes 2}\overset{p}\to\mathsf{E}[(1/\pi-1)g(D;\theta^\ast)^{\otimes 2}]$.

\noindent The second sets are regularity conditions for the underlying distribution of the data.

\noindent{\it C1}. \ $\Theta$ is compact and contains $\theta^\ast$ as an interior point.

\noindent{\it C2}. \ $R(\theta)$ is continuous in $\theta$ and $\theta^\ast$ is the unique global minimizer of $R(\theta)$ over $\Theta$.

\noindent{\it C3}. \ There exists a function $h(D)$ such that $\sup_{\theta\in\Theta}\|l(D;\theta)\|\leqslant h(D)$, $\sup_{\theta\in\Theta}\|g(D;\theta)\|\leqslant h(D)$,  $\sup_{\theta\in\Theta}\|G(D;\theta)\|\leqslant h(D)$, and $\mathsf{E}[h(D)]<\infty$. $G(D;\theta)$ is continuous in $\theta$ with probability one.

\noindent{\it C4}. \ The matrix $A$ is positive definite.

We now begin to prove Theorem \ref{thm1}. The following lemmas are needed.

\begin{lem}\label{lem1}
Under conditions A1, C1, and C3, $\sup_{\theta\in\Theta}|R_n(\theta)-R(\theta)|\overset{p}{\to}0$.
\end{lem}

\noindent{\it Proof:} \ By the triangle inequality, we have that
\begin{eqnarray}\label{l11}
& &\sup_{\theta\in\Theta}|R_n(\theta)-R(\theta)|\\
&\leqslant&\sup_{\theta\in\Theta}|R_n(\theta)-R_n^F(\theta)|+\sup_{\theta\in\Theta}|R_n^F(\theta)-R(\theta)|,\nonumber
\end{eqnarray}
where $R_n^F(\theta)=n^{-1}\sum_{i=1}^nl_i(\theta)$. By {\it C3} and uniform law of large number, it can be shown that $\sup_{\theta\in\Theta}\|R_n^F(\theta)-R(\theta)\|\overset{p}\to0$. Next we show the convergence of the first term in the right-hand-side of (\ref{l11}). For $i=1,\ldots,n$, define $d_\delta(\Delta_i,\theta_1)=\sup_{\theta\in B(\theta_1,\delta)}\Delta_il_i(\theta)/\pi_i-\inf_{\theta\in B(\theta_1,\delta)}\Delta_il_i(\theta)/\pi_i$, where $B(\theta_1,\delta)$ is a ball in $\mathbb{R}^p$ centered at $\theta_1$ with radius $\delta$. By the continuity of $l_i(\theta)$, the measurability of $l_i(\theta)/\pi_i$ to ${\cal F}_n$, and dominated convergence theorem, $\mathsf{E}[d_\delta(\Delta_i,\theta_1)|{\cal F}_n]\to0$ as $\delta\to0$. Thus, for all $\theta\in\Theta$ and $\varepsilon>0$, there exists $\delta_\varepsilon(\theta)>0$ such that $\mathsf{E}[d_{\delta_\varepsilon(\theta)}(\Delta_i,\theta_1)|{\cal F}_n]<\varepsilon$. Since $\Theta$ is compact, we can find a finite sequence of $\theta_1,\ldots,\theta_K$ such that $\Theta$ is covered by $\cup_{k=1}^KB(\theta_k,\delta_\varepsilon(\theta_k))$. Thus, we have that
\begin{eqnarray*}
& &\sup_{\theta\in\Theta}\left(R_n(\theta)-R_n^F(\theta)\right)\\
&=&\sup_{\theta\in\Theta}\frac{1}{n}\sum_{i=1}^n\left[\frac{\Delta_i}{\pi_i}l_i(\theta)
-\mathsf{E}\left(\frac{\Delta_i}{\pi_i}l_i(\theta)|{\cal F}_n\right)\right]\\
&\leqslant&\max_{1\leqslant k\leqslant K}\sup_{\theta\in B(\theta_k,\delta_\varepsilon(\theta_k))}\frac{1}{n}\sum_{i=1}^n\left[\frac{\Delta_i}{\pi_i}l_i(\theta)
-\mathsf{E}\left(\frac{\Delta_i}{\pi_i}l_i(\theta)|{\cal F}_n\right)\right]\\
&\leqslant&\max_{1\leqslant k\leqslant K}\frac{1}{n}\sum_{i=1}^n\left[\sup_{\theta\in B(\theta_k,\delta_\varepsilon(\theta_k))}\frac{\Delta_i}{\pi_i}l_i(\theta)-\mathsf{E}\left(\inf_{\theta\in B(\theta_k,\delta_\varepsilon(\theta_k))}\frac{\Delta_i}{\pi_i}l_i(\theta)|{\cal F}_n\right)\right].
\end{eqnarray*}
From {\it A1}, we have that $n^{-2}\sum_{i=1}^n(\sup_{\theta\in\Theta}l_i(\theta))^2/\pi_i\overset{p}\to0$. Thus, by Chebyshev inequality, {\it C3}, and the weak law of large number, we can show that
\begin{eqnarray*}
& &\max_{1\leqslant k\leqslant K}\frac{1}{n}\sum_{i=1}^n\left(\sup_{\theta\in B(\theta_k,\delta_\varepsilon(\theta_k))}\frac{\Delta_i}{\pi_i}l_i(\theta)
-\mathsf{E}\left(\inf_{\theta\in B(\theta_k,\delta_\varepsilon(\theta_k))}\frac{\Delta_i}{\pi_i}l_i(\theta)|{\cal F}_n\right)\right]\\
&=&\max_{1\leqslant k\leqslant K}\frac{1}{n}\sum_{i=1}^n\left[\mathsf{E}\left(\sup_{\theta\in B(\theta_k,\delta_\varepsilon(\theta_k))}\frac{\Delta_i}{\pi_i}l_i(\theta)|{\cal F}_n\right)-\mathsf{E}\left(\inf_{\theta\in B(\theta_k,\delta_\varepsilon(\theta_k))}\frac{\Delta_i}{\pi_i}l_i(\theta)|{\cal F}_n\right)\right]+a_n\\
&=&\max_{1\leqslant k\leqslant K}\frac{1}{n}\sum_{i=1}^n\mathsf{E}[d_{\delta_\varepsilon(\theta)}(\Delta_i,\theta_k)|{\cal F}_n]+a_n\leqslant\varepsilon+a_n,
\end{eqnarray*}
where $a_n$ satisfies that for all $\varepsilon_1>0$, $\mathsf{P}(|a_n|>\varepsilon_1|{\cal F}_n)\overset{p}\to0$. Similarly, we can show that $\inf_{\theta\in\Theta}(R_n(\theta)-R_n^F(\theta))\geqslant\varepsilon+b_n$, where for all $\varepsilon_2>0$, $\mathsf{P}(|b_n|>\varepsilon_2|{\cal F}_n)\overset{p}\to0$. Thus, we can show that for arbitrary $\varepsilon>0$, $\mathsf{P}(\sup_{\theta\in\Theta}|R_n(\theta)-R_n^F(\theta)|>\varepsilon|{\cal F}_n)\overset{p}\to0$. By Helly-Bray theorem, $\sup_{\theta\in\Theta}|R_n(\theta)-R_n^F(\theta)|\overset{p}\to0$. The desired conclusion follows from (\ref{l11}).\qed

\medskip

\begin{lem}\label{lem2}
Under condition A2 and C3, if $\tilde{\theta}\overset{p}{\to}\theta^\ast$, then $n^{-1}\sum_{i=1}^n\Delta_iG_i(\theta^\ast)/\pi_i\overset{p}{\to}A$.
\end{lem}

\noindent{\it Proof:} \ Conditioning on ${\cal F}_n$, $\Delta_iG_i(\theta^\ast)/\pi_i$, $i,=1,\ldots,n$, are independent with $\mathsf{E}[\Delta_iG_i(\theta^\ast)/\pi_i|{\cal F}_n]=G_i(\theta^\ast)$ and $\mathsf{E}[\|\Delta_iG_i(\theta^\ast)/\pi_i-G_i(\theta^\ast)\|^2|{\cal F}_n]=(1/\pi_i-1)\|G_i(\theta^\ast)\|^2$. By {\it A2}, Chebyshev inequality, {\it C3}, and the weak law of large number, for any $\varepsilon>0$, $\mathsf{P}(\|n^{-1}\sum_{i=1}^n\Delta_iG_i(\theta^\ast)/\pi_i-n^{-1}\sum_{i=1}^nG_i(\theta^\ast)\|>\varepsilon|{\cal F}_n)\overset{p}\to0$.
Then, by Helly-Bray theorem, we have that for any $\varepsilon>0$, $\mathsf{P}(\|n^{-1}\sum_{i=1}^n\Delta_iG_i(\theta^\ast)/\pi_i-n^{-1}\sum_{i=1}^nG_i(\theta^\ast)\|>\varepsilon)\to0$, that is
\begin{eqnarray}\label{l21}
\left\|\frac{1}{n}\sum_{i=1}^n\frac{\Delta_i}{\pi_i}G_i(\theta^\ast)-\frac{1}{n}\sum_{i=1}^nG_i(\theta^\ast)\right\|\overset{p}\to0.
\end{eqnarray}
By law of large number,
\begin{eqnarray}\label{l22}
\left\|\frac{1}{n}\sum_{i=1}^nG_i(\theta^\ast)-A\right\|\overset{p}\to0.
\end{eqnarray}
Thus, by the triangle inequality, (\ref{l21}), and (\ref{l22}), we have $n^{-1}\sum_{i=1}^n\Delta_iG_i(\theta^\ast)/\pi_i\overset{p}{\to}A$.\qed

\medskip

\begin{lem}\label{lem3}
Under condition C3, if $\tilde{\theta}\overset{p}{\to}\theta^\ast$, then $\tilde{A}\overset{p}{\to}A$.
\end{lem}

\noindent{\it Proof:} \ Define $A(\theta)=\mathsf{E}[G(D;\theta)]$. By the triangle inequality, we have that
\begin{eqnarray}\label{l31}
\|\tilde{A}-A\|\leqslant\|\tilde{A}-A(\tilde{\theta})\|+\|A(\tilde{\theta})-A\|.
\end{eqnarray}
Since $\tilde{\theta}\overset{p}\to\theta^\ast$, $\|\tilde{A}-A(\tilde{\theta})\|\leqslant\sup_{\theta\in{\cal N}(\theta^\ast)}\|n^{-1}\sum_{i=1}^nG_i(\theta)-A(\theta)\|$ as $n$ is sufficiently large, where ${\cal N}(\theta^\ast)$ is a neighborhood in $\theta$ about $\theta^\ast$. By {\it C3} and the uniform law of large number, $\sup_{\theta\in{\cal N}(\theta_0)}\|n^{-1}\sum_{i=1}^nG_i(\theta)-A(\theta)\|\overset{p}\to0$, so does $\|\tilde{A}-A(\tilde{\theta})\|$. By the continuity of $A(\theta)$ with respect to $\theta$, $\|A(\tilde{\theta})-A\|\overset{p}\to0$. Thus, by (\ref{l31}), $\tilde{A}\overset{p}\to A$.\qed

\medskip

\noindent{\it Proof of Theorem \ref{thm1}:} \ 1) Consistency: By {\it C2} and the uniform convergence of $R_n(\theta)$ to $R(\theta)$ on $\Theta$ from Lemma \ref{lem1}, we can show that $\hat{\theta}\overset{p}\to\theta^\ast$ by arguments similar to those in the proof of Theorem 5 in Fithian and Hastie (2014). The details are omitted here.

\noindent 2) Asymptotic normality: Since $\sum_{i=1}^n\Delta_ig_i(\hat{\theta})/\pi_i=0$, by Taylor expansion, we have that
\begin{eqnarray}\label{A11}
0&=&\frac{1}{\sqrt{n}}\sum_{i=1}^n\frac{\Delta_i}{\pi_i}g_i(\theta^\ast)+
\frac{1}{n}\sum_{i=1}^n\frac{\Delta_i}{\pi_i}G_i(\theta^\ast)\cdot\sqrt{n}\left(\hat{\theta}-\theta^\ast\right)\\
& &+o_p\left(\frac{1}{n}\sum_{i=1}^n\frac{\Delta_i}{\pi_i}G_i(\theta^\ast)\cdot\sqrt{n}\left(\hat{\theta}-\theta^\ast\right)\right).\nonumber
\end{eqnarray}

Write the first term in the right-hand-side of (\ref{A11}) as $\eta_{1n}+\eta_{0n}$, where
\begin{eqnarray*}
\eta_{1n}=\frac{1}{\sqrt{n}}\sum_{i=1}^n\frac{\Delta_i-\pi_i}{\pi_i}g_i(\theta^\ast)
\end{eqnarray*}
and $\eta_{0n}=n^{-1/2}\sum_{i=1}^ng_i(\theta^\ast)$. It is easy to see that conditioning on ${\cal F}_n$, $\eta_{1n}$ has zero mean and variance-covariance matrix equals to $n^{-1}\sum_{i=1}^n(1/\pi_i-1)g_i(\theta^\ast)^{\otimes 2}$. Form {\it A3}, the variance-covariance matrix converges in probability to $\mathsf{E}[(1/\pi-1)g(D;\theta^\ast)^{\otimes 2}]$ which is independent of ${\cal F}_n$. Therefore, $\eta_{1n}$ is asymptotically normal with mean zero and variance-covariance matrix $\mathsf{E}[(1/\pi-1)g(D;\theta^\ast)^{\otimes 2}]$. On the other hand, it is not difficult to see that $\mathsf{E}(\eta_{1n}+\eta_{0n}|{\cal F}_n)=\eta_{0n}$ which is asymptotically normal with mean zero and variance-covariance matrix $\mathsf{E}[g(D;\theta^\ast)^{\otimes2}]$. Consequently, $\eta_{1n}+\eta_{0n}$ is asymptotically normal with mean zero and variance-covariance matrix $\mathsf{E}[(1/\pi-1)g(D;\theta^\ast)^{\otimes 2}]+\mathsf{E}[g(D;\theta^\ast)^{\otimes2}]=V_\pi$.

Combining the asymptotic normality of $n^{-1/2}\sum_{i=1}^n\Delta_ig_i(\theta^\ast)/\pi_i$, Lemma \ref{lem2} and (\ref{A11}), we can derive that $\sqrt{n}(\hat{\theta}-\theta^\ast)=O_p(1)$ and
\begin{eqnarray*}
\sqrt{n}\left(\hat{\theta}-\theta^\ast\right)=-A^{-1}\frac{1}{\sqrt{n}}\sum_{i=1}^n\frac{\Delta_i}{\pi_i}g_i(\theta^\ast)+o_p(1),
\end{eqnarray*}
which is exactly (\ref{3.0}). Using Slutsky theorem, it follows quickly that $\sqrt{n}(\hat{\theta}-\theta^\ast)$ is asymptotically normal with mean zero and variance-covariance matrix $A^{-1}V_{\pi}A^{-1}$.\qed

\medskip

Next we turn to prove the corollaries. For Corollary \ref{cor1}, the following condition is needed.

\noindent{\it C5}. \ $\mathsf{E}(\sup_{\theta\in\Theta}\|l(D;\theta)\|^2/\pi)<\infty$ and $\mathsf{E}(\|G(D;\theta^\ast)\|^2/\pi)<\infty$, where $\pi$ is defined in Proposition \ref{pro1} or \ref{pro2}.

\noindent{\it Proof of Corollary \ref{cor1}:} \ The result can be proved from Theorem \ref{thm1} if one shows that {\it C1}-{\it C5} imply {\it A1}-{\it A3}. We take $\pi_i=(c\|\tilde{A}^{-1/2}g_i(\tilde{\theta})\|)\wedge1$ (for prediction accuracy) as an example. The proof of the other $\pi_i$ is of the same spirit. Note that $\pi_i$ can be regarded as the function of $\tilde{A}$, $\tilde{\theta}$, and the $i$th data point. We just write $\pi_i$ as $\pi(d_i;\tilde{A},\tilde{\theta})$, where $\pi(d_i;\tilde{A},\tilde{\theta})=(c\|\tilde{A}^{-1/2}g_i(\tilde{\theta})\|)\wedge1$. Correspondingly, write $\pi=(c\|A^{-1/2}g(D;\theta^\ast)\|)\wedge1$ as $\pi(D;A,\theta^\ast)$. By {\it C3}, it is easy to see that $\mathsf{E}(\|g(D;\theta^\ast)\|^2/\pi)<\infty$. Let $V_{1n}({\cal A},\theta)=n^{-1}\sum_{i=1}^n(1/\pi(d_i;{\cal A},\theta)-1)g_i(\theta^\ast)^{\otimes2}$, and $V_1({\cal A},\theta)=\mathsf{E}[(1/\pi(D;{\cal A},\theta)-1)g(D;\theta^\ast)^{\otimes2}]$, where ${\cal A}$ is a $p\times p$ matrix. By the triangle inequality, we have that
\begin{eqnarray}\label{A12}
& &\|V_{1n}(\tilde{A},\tilde{\theta})-V_1(A,\theta^\ast)\|\\
&\leqslant&\|V_{1n}(\tilde{A},\tilde{\theta})-V_1(\tilde{A},\tilde{\theta})\|
+\|V_1(\tilde{A},\tilde{\theta})-V_1(A,\theta^\ast)\|.\nonumber
\end{eqnarray}
Since $\tilde{\theta}\overset{p}\to\theta^\ast$ and $\tilde{A}\overset{p}\to A$ by Lemma \ref{lem3}, the first term in the right-hand-side of (\ref{A12}) is no larger than $\sup_{({\cal A},\theta)\in{\cal N}(A,\theta^\ast)}\|V_{1n}({\cal A},\theta)-V_{1}({\cal A},\theta)\|$ as $n$ is sufficiently large, where ${\cal N}(A,\theta^\ast)$ is a neighborhood in $({\cal A},\theta)$ about $(A,\theta^\ast)$. By {\it C3} and the uniform law of large number, $\sup_{({\cal A},\theta)\in{\cal N}(A,\theta^\ast)}\|V_{1n}({\cal A},\theta)-V_{1}({\cal A},\theta)\|\overset{p}\to0$. Moreover, $\|V_1(\tilde{A},\tilde{\theta})-V_1(A,\theta^\ast)\|\overset{p}\to0$ because of the continuity of $V_1({\cal A},\theta)$ with respect to $({\cal A},\theta)$. Thus, we have $V_{1n}(\tilde{A},\tilde{\theta})\overset{p}\to V_1(A,\theta^\ast)$, which means {\it A3} holds. By using similar arguments, we can show that  $n^{-1}\sum_{i=1}^n(\sup_{\theta\in\Theta}l_i(\theta))^2/\pi_i\overset{p}\to\mathsf{E}(\sup_{\theta\in\Theta}\|l(D;\theta)\|^2/\pi)$ and $n^{-1}\sum_{i=1}^n\|G_i(\theta^\ast)\|^2/\pi_i\overset{p}\to\mathsf{E}(\|G(D;\theta^\ast)\|^2/\pi)$. Then {\it C5} implies that {\it A1} and {\it A2} hold.\qed

\medskip

\noindent{\it Proof of Corollary \ref{cor2}:} \ When the covariates vector $X$ does not concentrate on a hyperplane of dimension smaller than $q$ and $\mathsf{E}(\|X\|^4/\pi)<\infty$, conditions {\it C3}-{\it C5} hold. When the Logistic model is correctly specified, it can be showed that $\Sigma_{\mbox{\footnotesize full}}=\{\mathsf{E}[p(\theta_0^\top Z)(1-p(\theta_0^\top Z))ZZ^\top]\}^{-1}$. From Theorem \ref{thm1}, when $\pi_i=|y_i-p(\tilde{\theta}^\top z_i)|$ and $\tilde{\theta}\overset{p}\to\theta_0$, $\sqrt{n}(\hat{\theta}-\theta_0)\overset{d}\to N(0,A^{-1}V_\pi A^{-1})$, where $A=\Sigma_{\mbox{\footnotesize full}}$ and in $V_\pi$, $g(Y,X;\theta)=(Y-p(\theta^\top Z))Z$ and $\pi=|Y-p(\theta_0^\top Z)|$. Thus, to obtain the conclusion, it is efficient to show that $V_\pi=2\Sigma_{\mbox{\footnotesize full}}^{-1}$. It is easy to see that
\begin{eqnarray*}
V_\pi=\mathsf{E}\left[\frac{g(Y,X;\theta_0)^{\otimes2}}{\pi}\right]=\mathsf{E}\left[\frac{(Y-p(\theta_0^\top Z))^2}{|Y-p(\theta_0^\top Z)|}ZZ^\top\right]=\mathsf{E}\left[|Y-p(\theta_0^\top Z)|ZZ^\top\right].
\end{eqnarray*}
For a 0-1 binary random variable $Y$ with $\mathsf{E}(Y)=p$, it is easy to see that $\mathsf{E}(|Y-p|)=2\mathsf{Var}(Y)=2p(1-p)$. Thus, we have that
\begin{eqnarray*}
& &\mathsf{E}\left[|Y-p(\theta_0^\top Z)|ZZ^\top\right]=\mathsf{E}\left\{E\left[|Y-p(\theta_0^\top Z)||Z\right]ZZ^\top\right\}\\
&=&2\mathsf{E}\left[p(\theta_0^\top Z)(1-p(\theta_0^\top Z))ZZ^\top\right]=2\Sigma_{\mbox{\footnotesize full}}^{-1}.
\end{eqnarray*}
The conclusion follows immediately.\qed

\medskip

To prove Corollary \ref{cor3}, we need another condition.

\noindent{\it C6}. \ $\bar{\theta}$ is an interior point of $\Theta$, the matrix $\bar{A}$ is positive definite, $\mathsf{E}(\sup_{\theta\in\Theta}\|l(D;\theta)\|^2/\bar{\pi})<\infty$, $\mathsf{E}(\|g(D;\theta^\ast)\|^2/\bar{\pi})<\infty$ and $\mathsf{E}(\|G(D;\theta^\ast)\|^2/\bar{\pi})<\infty$.

\noindent{\it Proof of Corollary \ref{cor3}:} \ We still take $\pi_i=(c\|\tilde{A}^{-1/2}g(d_i;\tilde{\theta})\|)\wedge1$ as an example.

\noindent 1) Consistency: The proof of consistency of $\hat{\theta}$ follows exactly the same step as that in Theorem \ref{thm1}, so we skip the details.

\noindent 2) Asymptotic normality: The proof of the asymptotic normality of $\hat{\theta}$ also follows the similar steps to those in Theorem \ref{thm1}. The main difference lies in that if $\tilde{\theta}\overset{p}\to\bar{\theta}$, by {\it C6}, we can show that $\tilde{A}\overset{p}\to\bar{A}$ using arguments similar to those in the proof of Lemma \ref{lem3}. Meanwhile, it can also be shown similarly that $V_{1n}(\tilde{A},\tilde{\theta})\overset{p}\to V_1(\bar{A},\bar{\theta})=V_{\bar{\pi}}$. Thus, $n^{-1/2}\sum_{i=1}^n\Delta_ig_i(\theta^\ast)/\pi_i$ is asymptotically normal with mean zero and variance-covariance matrix $V_{\bar{\pi}}$. Lemma \ref{lem2} still holds here. It follows then that $\sqrt{n}(\hat{\theta}-\theta^\ast)$ is asymptotically normal with mean zero and variance-covariance matrix $A^{-1}V_{\bar{\pi}}A^{-1}$.\qed

\bigskip

\noindent {\bf Acknowledgment}

The authors thank Professor Cheng Zhang and Pengfei Ma for providing the micro-blog data. The research of Wen Yu was supported by the National Natural Science Foundation of China Grants (11671097).

\medskip

\section{References}

\begin{enumerate}[{[}1{]}]
\item Ai, M., Yu, J., Zhang, H., Wang, H. (2018) Optimal subsampling algorithms for big data generalized linear models. {\it arXiv:1806.06761v1}.
\item Anderson, J. A. (1972). Separate sample logistic discrimination. {\it Biometrika} {\bf 59}, 19-35.
\item Breslow, N. E., Day, N. E. et al. (1980). {\it Statistical Methods in Cancer Research. The Analysis of Case-Control Studies 1.} Distributed for IARC by WHO, Geneva, Switzerland.
\item Branco, P., Torgo, L., and Ribeiro, P. R. (2015). A survey of predictive modelling under imbalanced distributions. {\it arXiv:1505.01658v2}.
\item Chen, K. (2001). Generalized case-cohort sampling. {\it Journal of the Royal Statistical Society {\rm B}}{\bf 63}, 791-809.
\item Chen, K. and Lo, S-H. (1999). Case-cohort and case-control analysis with Cox's model. {\it Biometrika} {\bf 86}, 755-764.
\item Fithian, W. and Hastie, T. (2014). Local case-control sampling: efficient subsampling in imbalanced data sets. {\it The Annals of Statistics} {\bf 42}, 1693-1724.
\item Horvitz, D. G. and Thompson, D. J. (1952). A generalized of sampling without replacement from a finite universe. {\it Journal of the American Statistical Association} {\bf 47}, 663-685.
\item Huber, P. J. (2011). {\it Robust Statistics}. Springer, Berlin.
\item Ma, P., Mahoney, M., and Yu, B. (2015). A statistical perspective on algorithmic leveraging. {\it Journal of Machine Learning Research} {\bf 16}, 861-911.
\item Mantel, N. and Haenszel, W. (1959). Statistical aspects of the analysis of data from retrospective studies of disease. {\it Journal of the National Cancer Institute} {\bf 22}, 719-748.
\item Miettinen, O. S. (1976). Estimability and estimation in case-referrent studies. {\it American Journal of Epidemiology} {\bf 104}, 226-235.
\item Prentice, R. L. (1986). A case-cohort design for epidemiologic cohort studies and disease prevention trials. {\it Biometrika} {\bf 73}, 1-11.
\item Prentice, R. L. and Pyke, R. (1979). Logistic disease incidence models and case-control studies. {\it Biometrika} {\bf 66}, 403-411.
\item Thomas, D. C. (1977). Appendum to ``Methods of cohort analysis: appraisal by application to asbestos mining," by  Liddell, F. D. K., McDonald, J. C. and Thomas, D. C. {\it Journal of the Royal Statistical Society {\rm A}} {\bf 140}, 469-490.
\item Wang, H. Yang, M., and Stufken, J. (2018). Information-based optimal subdata selection for big data linear regression. {\it Journal of the American Statistical Association} doi: 10.1080/01621459.2017.1408468.
\item Wang, H., Zhu, R., and Ma P. (2018). Optimal subsampling for large sample Logistic regression. {\it Journal of the American Statistical Association} doi: 10.1080/01621459.2017.1292914.
\item Webb, S., Caverlee, J., and Pu, C. (2006). Introducing the webb spam corpus: Using email spam to identify web spam automatically. In {\it Proceedings of the Third Conference on Email and Anti-Spam (CEAS)}. CEAS, Mountain View, CA.
\item Yao, Y., Yu, W., and Chen, K. (2017). End-point sampling. {\it Statistica Sinica}, {\bf 27}, 415-435.
\end{enumerate}

\end{document}